\documentclass[11pt]{article}

\usepackage{acl}

\usepackage{times}
\usepackage{latexsym}

\usepackage[T1]{fontenc}

\usepackage[utf8]{inputenc}

\usepackage{microtype}

\usepackage{inconsolata}

\usepackage{graphicx}

\usepackage{tikz}
\usepackage{algpseudocode}
\algrenewcommand\textproc{\text}
\usepackage{makecell}
\usepackage{booktabs}
\usepackage{color}
\usepackage{colortbl}
\usepackage{multirow}
\usepackage{enumitem}
\usepackage{makecell}
\usepackage{threeparttable}
\usepackage{amsmath,amsfonts,mathtools}
\usepackage{pifont}

\usepackage{float}
\usepackage{graphicx}
\usepackage{xcolor}
\usepackage{enumitem}
\usepackage{tabularx, booktabs}

\usepackage{tcolorbox}

\usepackage{colortbl}
\usepackage{float}
\usepackage{graphicx}
\usepackage{hyperref}
\usepackage{tabularx, booktabs}
\usepackage{tikz}
\usepackage{CJKutf8}

\usepackage{bm}        
\usepackage{amsfonts}
\usepackage{cancel}

\definecolor{mygreen}{HTML}{00B050}

\definecolor{myorange}{HTML}{ED7D31}
\definecolor{rowgray}{gray}{0.97}
\definecolor{avgblue}{RGB}{210,230,250}  
\definecolor{headergray}{RGB}{160,160,160} 
\definecolor{uc_color}{rgb}{0.99,0.24,0.63}
\definecolor{hc_color}{rgb}{0.02,0.51,0.51}
\definecolor{tc_color}{rgb}{0.99,0.55,0.09}
\definecolor{posgreen}{RGB}{0,150,0}
\definecolor{negred}{RGB}{200,0,0}

\usepackage{array}
\usepackage{mathtools}
\usepackage{multirow}
\usepackage{subcaption}
\usepackage{tikz}
\renewcommand{\arraystretch}{1.1}
\definecolor{color1}{cmyk}{0.216,0.176,0,0}
\definecolor{color2}{cmyk}{0.059,0.235,0.392,0}

\usepackage{graphicx}
\usepackage{tikz}
\usepackage{wrapfig}
\usepackage{algorithm}
\usepackage{algpseudocode}
\algrenewcommand\textproc{\text}
\usepackage{makecell}
\usepackage{booktabs}
\usepackage{pifont}
\usepackage{multirow}
\usepackage{enumitem}
\usepackage{balance}
\usepackage{threeparttable}
\usepackage{amsmath,amsfonts,mathtools} 
\usepackage{longtable}
\usepackage{amsthm}

\tikzstyle{mybox} = [draw=black, very thick,
    rectangle, rounded corners, inner sep=10pt, inner ysep=13pt]
\tikzstyle{fancytitle} =[fill=black, text=white]

\newcommand{\M}{\textsc{Harness}-RL}

\title{\textsc{Harness}-RL: Black-Box Reinforcement Learning with Action--Args Decoupling for Central-Agent Multi-Agent Harnesses}
\author{
\textbf{
Xinke Jiang\textsuperscript{1,2,3}\thanks{Equal contribution.},
Zhixin Zhang\textsuperscript{1,2,3,*},
Zhibang Yang\textsuperscript{1,2,3,*},
Jiaran Gao\textsuperscript{1,2,3},
Rihong Qiu\textsuperscript{1,2,3},
}
\\[-0.1em]
\textbf{
Shijin Chen\textsuperscript{4},
Xu Chu\textsuperscript{2,3,5}\thanks{Corresponding authors.},
Junfeng Zhao\textsuperscript{2,3,$\dagger$},
Yasha Wang\textsuperscript{1,6,$\dagger$}
}
\\[-0.2em]
\normalfont
\textsuperscript{1}National Engineering Research Center of Software Engineering, Peking University, Beijing, China\\[-0.2em]
\textsuperscript{2}School of Computer Science, Peking University, Beijing, China\\[-0.2em]
\textsuperscript{3}Key Laboratory of High Confidence Software Technologies, Ministry of Education, Beijing, China\\[-0.2em]
\textsuperscript{4}Guangxi Land and Resources Planning and Design Group Co., Ltd, Guangxi, China\\[-0.2em]
\textsuperscript{5}Center on Frontiers of Computing Studies, Peking University, Beijing, China\\[-0.2em]
\textsuperscript{6}Peking University Information Technology Institute (Tianjin Binhai), Tianjin, China\\[-0.2em]
{\small
\{xinkejiang, yangzb\}@stu.pku.edu.cn,
\{chu\_xu, zhaojf, wangyasha\}@pku.edu.cn
}
}

\begin{document}
\maketitle
\begin{abstract}
Large language model agents increasingly solve long-horizon tasks through multi-agent harnesses in which a central agent coordinates specialized sub-agents, tools, and environments. Training the central policy in such a harness raises two challenges. First, an action label is a low-cardinality decision, whereas its args form a high-dimensional conditional sequence; optimizing both with a shared sequence-level signal can produce conflicting gradients. Second, dynamic scheduling creates interdependent sessions with branches, parallel calls, and rewritten contexts, which cannot be faithfully reduced to one flat token sequence. We introduce \M, a structured reinforcement learning framework that combines Conflict-Aware Policy Optimization (CAPO) with interface-level black-box trajectory construction. 
The black-box component captures Interface Call Records, builds per-session prefix trees, and aligns outcome and process rewards with trainable tokens.
CAPO uses forward activations to identify parameter partitions associated with action and args tokens, then routes their policy gradients to the corresponding subspaces. \M~supports both central-only and joint multi-agent training. Across seven multi-hop question answering and agentic retrieval benchmarks, it reaches average F1 scores of 42.93 and 47.79 with Qwen2.5-1.5B and Qwen2.5-3B, respectively, while ablations validate the contribution of CAPO and favor central-only optimization in the evaluated setting. Our code is available at 
{\url{https://github.com/jiangxinke/Harness-RL}}.
\end{abstract}

\section{Introduction}
\textbf{Large Language Models (LLMs)} continue to advance~\cite{guo2025deepseek, openaio1,yang2024qwen2}, their applications are expanding beyond single-turn text generation toward agent systems capable of planning~\cite{yao2022react}, tool use~\cite{jiang2024tcrag}, and interaction with external environments.
For complex reasoning, long-horizon task execution, and decision making in open environments, a single agent that generates a linear sequence is often insufficient. Researchers have therefore developed sophisticated agent harnesses that organize a central agent~\cite{DBLP:journals/corr/abs-2406-13381, DBLP:conf/acl/YueZLWWCQ25,zhang2026stackplanner}, specialized sub-agents~\cite{DBLP:journals/corr/abs-2308-08155,DBLP:conf/iclr/HongZCZCWZWYLZR24, DBLP:conf/acl/QianLLCDL0CSCXL24}, external tools, memory modules, and an execution environment to decompose, execute, and coordinate complex tasks~\cite{DBLP:conf/ijcai/GuoCWCPCW024, chen2025surveyllmbasedmultiagentsystem}.
\begin{figure*}[t]
  \centering
  \includegraphics[width=0.8\linewidth]{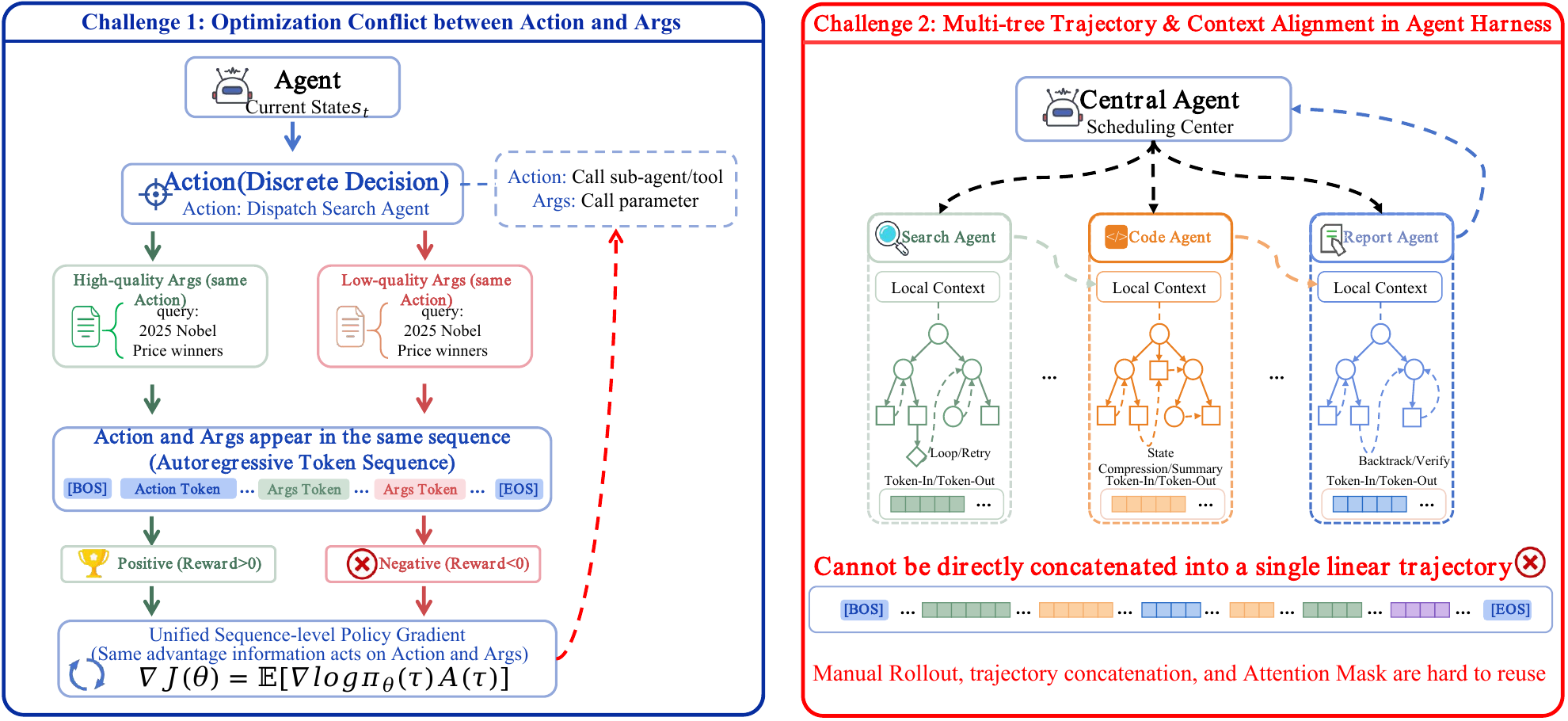}
\caption{Two challenges addressed by \M. Left: unified
  sequence-level optimization entangles Action and Args gradients. Right:
  dynamic multi-agent execution produces context-dependent trajectory trees that
  cannot be directly concatenated.}
  \label{fig:motivation}
  \vspace{-0.5cm}
\end{figure*}

Within such systems, the harness organizes the overall task workflow, while the central agent generally serves as the core policy model. \textbf{The central agent is responsible for global task decomposition, sub-task dispatch, context maintenance, and multi-round decision coordination. Specialized sub-agents, by contrast, perform concrete operations within their respective capability domains,} such as retrieval~\cite{jin2025search, trivedi2023interleaving}, code execution~\cite{DBLP:journals/corr/abs-2308-08155, DBLP:conf/acl/QianLLCDL0CSCXL24}, and tool invocation~\cite{jiang2024tcrag}. Every decision made by the central agent affects the subsequent execution path and environment state. An incorrect action or a low-quality args can therefore have a cumulative effect over a long interaction. \textbf{Consequently, this raises a key question: how can we effectively train the central agent's policy within a multi-agent harness to improve overall system performance?}

Existing studies have made substantial progress in multi-agent system architectures~\cite{DBLP:journals/corr/abs-2308-08155,DBLP:conf/iclr/HongZCZCWZWYLZR24, DBLP:conf/acl/QianLLCDL0CSCXL24}, communication mechanisms~\cite{DBLP:conf/icml/Du00TM24, DBLP:conf/nips/YangCSSQRZ25}, workflow design~\cite{DBLP:conf/iclr/QianXW0ZXDDC00025,DBLP:conf/iclr/ZhangXYTCCZCHWZ25,DBLP:conf/nips/HuZFNYXSJLZWYGL25}, and memory management~\cite{zhang2026stackplanner}. Meanwhile, recent work such as Search-R1~\cite{jin2025search} has demonstrated the effectiveness of reinforcement learning (RL) for tool use and agentic task execution. However, existing agentic RL methods generally make two implicit assumptions. At the policy-optimization level, different types of generated tokens can be jointly updated under a unified sequence-level objective. At the trajectory-modeling level, interactions within an agent harness can be reconstructed as a single linear token sequence according to a predefined workflow. 
As illustrated in Figure~\ref{fig:motivation}, these assumptions no longer hold in agent harnesses with dynamic sub-agent scheduling, conditional branches, loops, concurrency, backtracking, and context compression, giving rise to two challenges in policy optimization and trajectory modeling:

\noindent\textit{\textbf{\ding{182} Challenge 1: optimization conflicts between heterogeneous Action and Args decisions.}}
Agents at different levels make decisions through Action--Args pairs. A central agent selects an operation or dispatches a sub-agent through an Action, while a sub-agent uses an Action to execute a task or invoke a tool. Args describe the corresponding sub-task content or invocation args. An Action is a low-cardinality discrete decision, whereas Args constitute high-flexibility sequence generation conditioned on that Action. Although both occur in the same autoregressive sequence, they differ in decision characteristics and optimization objectives. For example, the same \textsc{Dispatch} Action may be paired with either a high-quality search query or an irrelevant or erroneous one. In this case, the difference in trajectory reward is primarily caused by the quality of Args. Standard GRPO~\cite{shao2024deepseekmath} or DAPO~\cite{DBLP:conf/nips/YuZZYZYDFLLLLLL25}, however, computes policy gradients uniformly over the full sequence, allowing the negative update induced by poor Args to interfere with an otherwise correct Action decision. As a result, the model may exhibit both unstable action planning and inadequate args generation.

\noindent\textit{\textbf{\ding{183} Challenge 2: multi-tree trajectory modeling and context alignment in a multi-agent harness.}}
The central agent dynamically schedules multiple sub-agents, whose executions may themselves contain conditional branches, loops, concurrency, backtracking, and context compression. The complete harness interaction therefore no longer corresponds to a single trajectory tree; instead, it forms multiple related agent interaction trees. Contexts associated with different agents and branches must remain isolated, while retaining the cross-agent dependencies created by task dispatch, message passing, and shared environment state. These trajectories cannot be concatenated directly. \textbf{Manually implementing rollout scripts, trajectory concatenation, and token-mask logic for each harness incurs substantial training-system overhead and is difficult to reuse.}

To address these challenges, we propose \M, a black-box reinforcement learning framework with action--args decoupling for central-agent-driven multi-agent harnesses. The framework consists of two connected components: Conflict-Aware Policy Optimization (CAPO) and a black-box RL mechanism for multi-agent harnesses. \textbf{CAPO constructs functional partitions for Action and Args and decouples their gradients. The black-box RL mechanism converts actual harness interactions into context-aligned training trajectories with attributable rewards. }Together, the two components provide a unified solution spanning heterogeneous-decision optimization and complex multi-agent trajectory modeling. \textbf{\ding{182} For \textbf{\textit{C1}}, we introduce CAPO.} It uses activation signals from model forward passes to measure the response strength of Action Tokens and Args Tokens across layers and structural units, and uses these statistics to construct the corresponding functional parameter partitions. During RL optimization, policy gradients produced by Action Tokens and Args Tokens primarily update their respective partitions. This design decouples action selection from args generation and reduces parameter interference between the two heterogeneous decision types. \textbf{\ding{183} For \textbf{\textit{C2}}, we introduce a black-box RL mechanism for multi-agent harnesses.} The mechanism treats the harness as a black box and captures the input and output of each agent interaction. It reconstructs multiple related interaction trees according to inter-agent invocation relationships, thereby representing dynamic scheduling, branched execution, and cross-agent contextual dependencies in a unified form. It further groups contiguous generated content into action-level decision units and attributes global outcome rewards and local process rewards to the relevant agents and decisions. This enables trajectory modeling, context alignment, reward attribution, and policy optimization for complex multi-agent interactions. Our main contributions are as follows:

\begin{itemize}[leftmargin=*,noitemsep]
      \item We formulate policy training for central-agent-driven multi-agent harnesses and propose \M, which converts long-horizon, multi-round interactions among agents, tools, and the environment into structured RL trajectories and supports both central-only and joint multi-agent training.

      \item We combine CAPO, which uses forward activations and token-specific gradient routing to decouple Action and Args optimization, with a black-box RL mechanism that reconstructs multi-session trajectories and aligns outcome and process rewards with trainable decisions.

      \item Experiments on multiple multi-agent reasoning and collaboration benchmarks show that \M~improves planning, dispatch, and coordination while outperforming strong baselines. Ablations further validate CAPO and the choice of training scope.
  \end{itemize}

\section{Preliminaries}
\label{sec:preliminaries}

\subsection{Central-Agent Multi-Agent Harnesses}

We consider a structured multi-agent harness whose policy layer contains a central agent and a set of specialized sub-agents. Let $\mathcal{R}=\{\mathsf{c}\}\cup\mathcal{A}, \mathcal{A}=\{\mathsf{a}_1,\ldots,\mathsf{a}_N\},$
where $\mathsf{c}$ denotes the central agent and $\mathcal{A}$ the sub-agent set.

The harness $\mathcal{H}$ is an execution and control layer outside the policies. It manages messages, workflow control, dispatch, tool invocation, state updates, and termination. Tools, search engines, and task environments are execution resources exposed through $\mathcal{H}$. Given a task $q$, the harness coordinates multiple rounds of interaction until termination. We call the complete interaction a \emph{harness rollout}. The
$e$-th rollout is
\begin{equation}
\tau_e = \left\{
\left(r_{e,j},s_{e,j},a_{e,j}^{(r_{e,j})},o_{e,j+1}\right)
\right\}_{j=0}^{T_e-1},
\label{eq:rollout}
\end{equation}
where $j$ indexes interaction steps, $r_{e,j}\in\mathcal{R}$ is the scheduled
role, $s_{e,j}$ is that role's visible state,
$a_{e,j}^{(r_{e,j})}$ is its action, and $o_{e,j+1}$ is the resulting
observation, such as a sub-agent response, tool result, or environment feedback.
Local contexts may be isolated across roles, while dispatch, messaging, and
execution state connect them at the rollout level.

\paragraph{Central agent.}
The central agent interprets the task, constructs and revises a plan, dispatches
sub-tasks, and maintains global state. For long-horizon execution, it maintains
a dynamic memory $M_j^{(\mathsf{c})}$:
\begin{equation}
M_{j+1}^{(\mathsf{c})}=\operatorname{Update}_{\mathsf{c}}
\left(M_j^{(\mathsf{c})},a_j^{(\mathsf{c})},o_{j+1}^{(\mathsf{c})}\right).
\end{equation}
We represent its structured decision as
\begin{equation}
a_j^{(\mathsf{c})}=\left(u_j^{(\mathsf{c})},z_j^{(\mathsf{c})}\right),
\label{eq:action_args}
\end{equation}
where $u_j^{(\mathsf{c})}\in \{\textsc{Reason},\textsc{Dispatch},\textsc{Summary}\},$ is the action and $z_j^{(\mathsf{c})}$ is args sequence. \textsc{Reason} represents planning and decision analysis; \textsc{Dispatch} specifies a sub-agent task and its args; and \textsc{Summary} consolidates intermediate results and may support revision.

\paragraph{Sub-agents.}
A sub-agent $\mathsf{a}_i$ executes an assigned task in a isolated context and may invoke tools, search engines, or the task environment. Its result is returned to the harness, serialized as an observation for the central agent, and incorporated into subsequent state. In the central-only setting, sub-agents are part of the environment and receive no policy gradient. In joint training, sub-agents are trainable policies with their own sessions.

\subsection{Policy Optimization in a Multi-Agent Harness}

Let $\mathcal{R}_{\mathrm{train}}\subseteq\mathcal{R}$ denote the trainable
roles. The central-only setting uses
$\mathcal{R}_{\mathrm{train}}=\{\mathsf{c}\}$; joint training may include sub-agents. For role $r$, let $\theta_r$ be its policy parameters and define $\Theta_{\mathrm{train}}=\{\theta_r\mid r\in\mathcal{R}_{\mathrm{train}}\}.$ At interaction step $j$, the scheduled role $r_j$ samples an action
$a_j^{(r_j)}$ from its policy conditioned on the visible state $s_j$. The harness then executes the action, returns observation $o_{j+1}$, and updates the interaction state to $s_{j+1}$. Together, the trainable role policies and the harness induce the rollout distribution $\Pi_{\Theta_{\mathrm{train}},\mathcal{H}}(\cdot\mid q)$. With rollout reward $R(\tau)$, the learning objective is
\begin{equation}
\Theta_{\mathrm{train}}^*=
\arg\max_{\Theta_{\mathrm{train}}}
\mathbb{E}_{q\sim\mathcal{D},\,
\tau\sim\Pi_{\Theta_{\mathrm{train}},\mathcal{H}}(\cdot\mid q)}
\left[R(\tau)\right].
\label{eq:multiagent_objective}
\end{equation}

Harness rollouts differ from ordinary prompt--response optimization in three ways: they are long-horizon, combine heterogeneous decisions, and continually condition on dynamic observations. Standard GRPO or DAPO typically broadcasts one rollout advantage to all trainable response tokens. This can couple action selection with args generation. Moreover, multiple model calls and role switches do not naturally form one trainable token sequence. Section \ref{sec:methodology} addresses these optimization and trajectory-construction problems jointly.

\section{\M}\label{sec:methodology}
This section presents the core methodology of \M. Section~\ref{sec:framework} provides an overview of the framework; Section~\ref{sec:capo} introduces Conflict-Aware Policy Optimization (CAPO); and Section~\ref{sec:blackbox} presents a black-box reinforcement learning method for multi-agent harnesses.

\subsection{Framework Overview}
\label{sec:framework}
\begin{figure*}
    \centering
    \includegraphics[width=0.9\textwidth]{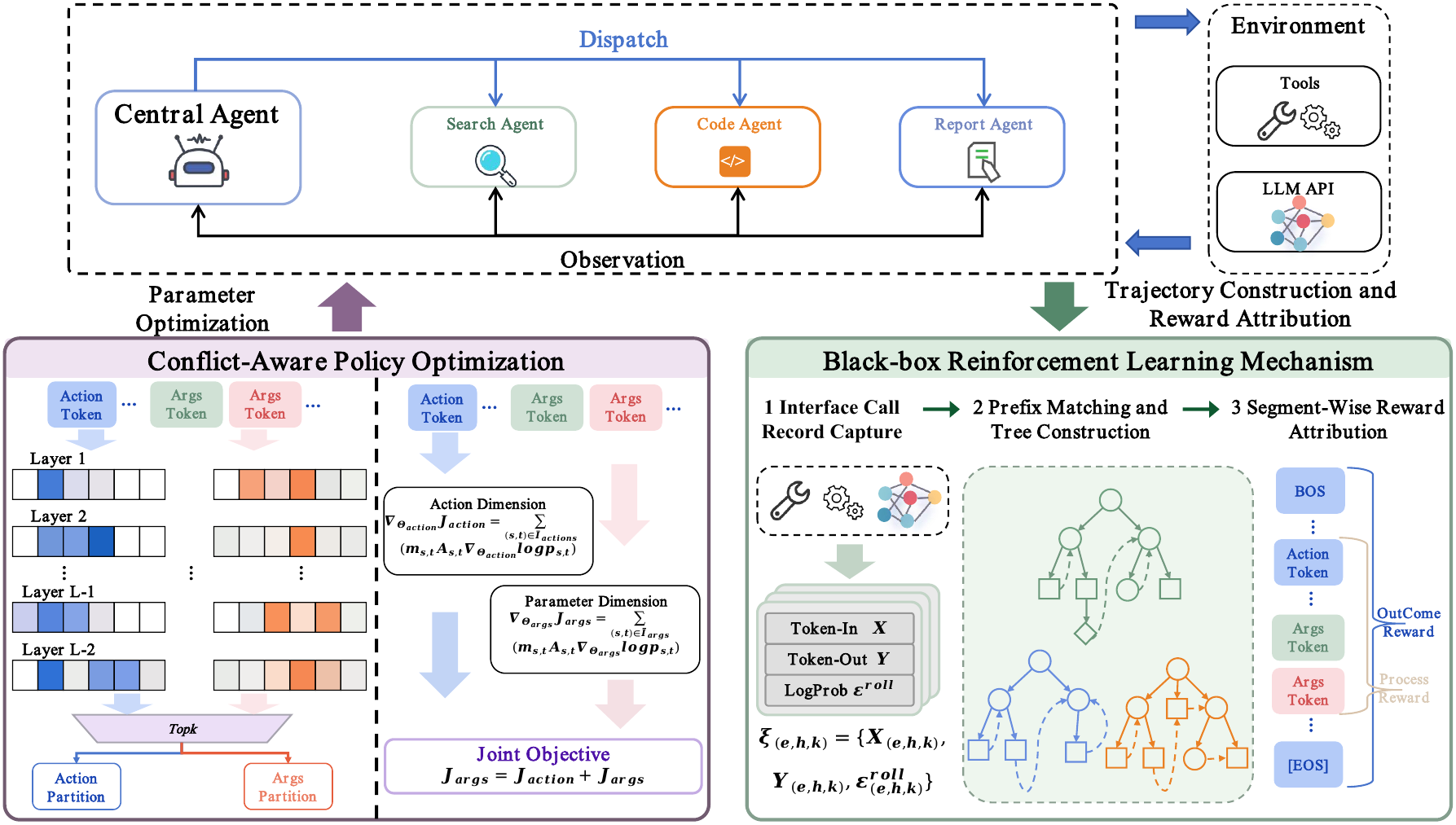}
    \caption{Overall Framework of \M.}
    \label{fig:newmethod.png}
    \vspace{-0.5cm}
\end{figure*}
\M~couples CAPO with an interface-level black-box RL mechanism. The black-box mechanism converts harness interactions into token-aligned training data, while CAPO routes the gradients induced by Action and Args tokens to their respective functional parameter subsets. For each query $q$, the two components operate on a shared structured training set $\mathcal{D}_{\mathcal{H}}(q)$, abbreviated as $\mathcal{D}_{\mathcal{H}}$ when the query is clear. On this dataset, CAPO decomposes the policy objective into Action and Args terms: 

\begin{equation}
J_{\text{\M}}(\mathcal{D}_{\mathcal{H}}) =
J_{\mathrm{action}}(\mathcal{D}_{\mathcal{H}})+
J_{\mathrm{args}}(\mathcal{D}_{\mathcal{H}}).
\label{eq:framework_objective}
\end{equation}
Here, $J_{\mathrm{action}}$ and $J_{\mathrm{args}}$ denote the contributions from the corresponding token types; Section~\ref{sec:capo} details their optimization.
The construction of $\mathcal{D}_{\mathcal{H}}$ proceeds in three stages. First, at call $k$ of session $h$ in rollout $\tau_e$, the black-box mechanism captures an Interface Call Record $\xi_{e,h,k}$ containing the exact input tokens, output tokens, and aligned sampling log-probabilities. Second, prefix matching organizes the records within each session into a token-prefix tree, from which captured output spans directly yield base samples in $\widetilde{\mathcal{D}}_{\mathcal{H}}(\tau_e)$. Third, rollout-group outcome and process rewards are converted into token-aligned advantages, producing the structured training set $\mathcal{D}_{\mathcal{H}}$. Section~\ref{sec:blackbox} formalizes these three stages.

\subsection{Conflict-Aware Policy Optimization}
\label{sec:capo}

\paragraph{Conflict between Heterogeneous Decisions}
Let $\mathcal{I}_{\mathrm{action}}$ and $\mathcal{I}_{\mathrm{args}}$ denote action- and args-token positions. For a trainable role $r\in\mathcal{R}_{\mathrm{train}}$, we write $\theta=\theta_r$ when the role is clear; CAPO is applied independently to each trainable role. Standard
sequence-level optimization can be written as
\begin{equation}
J(\theta)=J_{\mathrm{action}}(\theta)+J_{\mathrm{args}}(\theta),
\end{equation}
\begin{equation}
\nabla_\theta J=
\nabla_\theta J_{\mathrm{action}}+
\nabla_\theta J_{\mathrm{args}}.
\end{equation}
The two decisions have different functions: action tokens select high-level operations, while args tokens specify executable content. Their updates conflict when $\left\langle \nabla_\theta J_{\mathrm{action}}, \nabla_\theta J_{\mathrm{args}} \right\rangle < 0.$ Applying both gradients to the full parameter space may therefore impair action selection, args quality, or both.

CAPO constructs the functional partitions and applies gradient routing in three
steps.

\paragraph{\ding{182} High-Quality Trajectory Construction.} For every training query, CAPO samples multiple agentic rollouts and retains successful ones as a probing set. This rejection-sampling procedure reduces contamination from activation patterns associated with failed behavior.

\paragraph{\ding{183} Activation Statistics and Functional Partitioning.}
Assume the policy has $L$ structural units and let $h_{l,j,t}^{(n)}$ denote the activation of unit $j$ in layer $l$ at token $t$
for probing example $n$. We estimate decision-specific importance by
\begin{align}
I_{\mathrm{action}}(l,j)
&=\frac{1}{N}\sum_{n=1}^{N}
\frac{1}{|\mathcal{I}_{\mathrm{action}}^{(n)}|}
\sum_{t\in\mathcal{I}_{\mathrm{action}}^{(n)}}
\phi\left(h_{l,j,t}^{(n)}\right), \\
I_{\mathrm{args}}(l,j)
&=\frac{1}{N}\sum_{n=1}^{N}
\frac{1}{|\mathcal{I}_{\mathrm{args}}^{(n)}|}
\sum_{t\in\mathcal{I}_{\mathrm{args}}^{(n)}}
\phi\left(h_{l,j,t}^{(n)}\right),
\label{eq:activation_importance}
\end{align}
where $\phi(\cdot)$ measures positive activation strength. We select the most responsive units for each behavior:
$\mathcal{U}_{\mathrm{action}}=\operatorname{TopK} \left(I_{\mathrm{action}},K_{\mathrm{action}}\right),$
and $\mathcal{U}_{\mathrm{args}}=\operatorname{TopK} \left(I_{\mathrm{args}},K_{\mathrm{args}}\right).$ These unit sets induce parameter- or unit-level gradient masks. Units selected by both sets are treated as shared and follow the implementation's overlap rule. For the role under optimization, $\theta_{\mathrm{action}},\theta_{\mathrm{args}}\subseteq\theta_r$ denote the parameters associated with the selected units.

\paragraph{\ding{184} Token-Specific Gradient Routing.}

For role $r$, let $\mathcal{S}_e^{(r)}\subseteq \widetilde{\mathcal{D}}_{\mathcal{H}}(\tau_e)$ be its linear samples from rollout $e$. For sample $S_n$, let $\mathcal{I}_{\mathrm{action},n}$ and $\mathcal{I}_{\mathrm{args},n}$ be disjoint trainable response positions $\mathcal{I}_{\mathrm{action},n}\cap \mathcal{I}_{\mathrm{args},n}=\emptyset$ and $\mathcal{I}_{\mathrm{action},n}\cup \mathcal{I}_{\mathrm{args},n} \subseteq\mathcal{I}_{\mathrm{out},n}.$

Action spans are identified from structured action markers; args spans are identified from the corresponding fields. Inputs, observations, replayed context, and frozen-role outputs are excluded. Aggregate positions as
\begin{equation}
\overline{\mathcal{I}}_b^{(r)}=
\{(n,t)\mid \exists e,\;S_n\in\mathcal{S}_e^{(r)},\;
t\in\mathcal{I}_{b,n}\}.
\end{equation}
Where $b\in\{\mathrm{action},\mathrm{args}\}$. Let $P_n$ be the prompt length and $y_{n,t}=x_{n,P_n+t}$ the $t$-th response token. Define its policy log-probability as
\begin{equation}
\ell_{n,t}(\theta)=
\log\pi_\theta(y_{n,t}\mid x_{n,\leq P_n+t-1}).
\end{equation}
For $b\in\{\mathrm{action},\mathrm{args}\}$, and omitting clipping for clarity, the routed gradient of the corresponding term in
Eq.~\eqref{eq:framework_objective} is
\begin{equation}
\nabla_{\theta_b}J_b
=\sum_{(n,t)\in\overline{\mathcal{I}}_b^{(r)}}
A_{n,t}\nabla_{\theta_b}\ell_{n,t}(\theta).
\label{eq:routed_gradients}
\end{equation}
By construction, $\overline{\mathcal{I}}_b^{(r)}$ contains only trainable positions; $A_{n,t}$ is the token-aligned advantage defined in Eq.~\eqref{eq:token_advantage}.

\subsection{Interface-Level Black-Box RL}
\label{sec:blackbox}

Because role-specific calls maintain isolated contexts and interact through harness messages and observations, a harness rollout is not a single linear token sequence. We construct trainable RL data in three steps: Interface Call Record capture, session-wise token-prefix tree construction, and rollout-group reward attribution.

\paragraph{\ding{182} Interface Call Records.}

For session $h$ in rollout $e$, let $r(h)$ be its role. Before call $k$, the harness provides messages $\mathcal{C}_{e,h,k}$ and tool specifications $U_{e,h,k}$. Applying the model's chat template gives the exact Token-In:
\begin{equation}
\mathbf{X}_{e,h,k}=\operatorname{Template}
(\mathcal{C}_{e,h,k},U_{e,h,k}).
\end{equation}
The role policy samples Token-Out and stores aligned rollout log-probabilities:
\begin{align}
\mathbf{Y}_{e,h,k}&\sim
\pi_{\theta_{r(h)}}^{(r(h))}(\cdot\mid\mathbf{X}_{e,h,k}),\\
\boldsymbol{\ell}_{e,h,k}^{\mathrm{roll}}
&=(\ell_{e,h,k,1}^{\mathrm{roll}},\ldots,
\ell_{e,h,k,|\mathbf{Y}_{e,h,k}|}^{\mathrm{roll}}).
\end{align}
Thus one call record is
$\xi_{e,h,k}=\left(\mathbf{X}_{e,h,k},\mathbf{Y}_{e,h,k},\boldsymbol{\ell}_{e,h,k}^{\mathrm{roll}}\right),$
with rollout, session, role, and call identifiers stored as metadata. This record contains only model-side training information. The harness executes the output and writes tool, sub-agent, or environment feedback into the next
message context:$ (\sigma_{e,h,k+1},\mathcal{C}_{e,h,k+1})= \operatorname{Transition}_{\mathcal{H}} (\sigma_{e,h,k},\mathcal{C}_{e,h,k},\mathbf{Y}_{e,h,k}).$
Here, $\sigma_{e,h,k}$ is the harness execution state. The transition connects the captured segment to the context of subsequent calls.

\paragraph{\ding{183} Prefix Matching and Tree Construction.}
Let $\mathcal{U}_e$ be the sessions in rollout $e$. For each captured call,
concatenate its exact input and output tokens as
$\mathbf{Z}_{e,h,k}=\mathbf{X}_{e,h,k}\oplus\mathbf{Y}_{e,h,k}$. Calls in
each session form a prefix tree, and the session trees form the rollout-level
tree collection:
\begin{equation}
\begin{aligned}
\mathcal{T}_{e,h}
&=\operatorname{PrefixTree}
\big(\{\mathbf{Z}_{e,h,k}\}_{k=1}^{K_{e,h}}\big),\\
\mathcal{T}_e
&=\{\mathcal{T}_{e,h}\}_{h\in\mathcal{U}_e}.
\end{aligned}
\label{eq:prefix_trees}
\end{equation}
Shared prefixes are merged and unmatched suffixes form separate branches. A
rewritten context is attached at its longest shared prefix, while the
pre-rewrite output remains on the original branch. Thus $\mathcal{T}_e$
encodes token visibility rather than causal execution. Each output-span node $v(n)$ in $\mathcal{T}_e$ corresponds to one Interface Call Record $\xi_n$ and defines one base sample:
\begin{equation}
\begin{aligned}
S_n&=(\xi_n,\mathbf{m}_n,\eta_n),\\
\widetilde{\mathcal{D}}_{\mathcal{H}}(\tau_e)
&=\{S_n\mid v(n)\in\mathcal{V}_{\mathrm{out}}(\mathcal{T}_e)\}.
\end{aligned}
\label{eq:interface_samples}
\end{equation}
Here $\mathcal{V}_{\mathrm{out}}(\mathcal{T}_e)$ denotes captured output-span
nodes. The record $\xi_n$ supplies the exact input tokens, output tokens, and
aligned sampling log-probabilities. The mask satisfies $m_{n,t}=1$ only for new
Token-Out positions from trainable roles and is zero elsewhere, while $\eta_n$
stores call and tree-node metadata. For $n=(e,h,k)$, the response length is
$L_n^{\mathrm{resp}}=|\mathbf{Y}_{e,h,k}|$. Reusing the captured token IDs
avoids retokenization mismatch.

\paragraph{\ding{184} Segment-Wise Reward Attribution.}
For query $q$, sample a rollout group $G(q)=\{\tau_e\}_{e=1}^{K}$. The outcome reward and its group-normalized advantage are
\begin{align}
R_e^{\mathrm{out}}
&=\operatorname{F1}_{\mathrm{token}}(y_e,y_e^*)
I_e^{\mathrm{artifact}},\\
A_e^{\mathrm{out}}
&=\frac{R_e^{\mathrm{out}}-\mu_{G(q)}^{\mathrm{out}}}
{\sigma_{G(q)}^{\mathrm{out}}+\delta},
\end{align}
where $I_e^{\mathrm{artifact}}$ indicates whether the required output constraints are satisfied. We parse trainable Action--Args spans into decision units $\mathcal{Q}_e$. For unit $d$, let $\mathcal{I}_{e,d}$ contain its trainable token positions. Its process reward and normalized advantage are
\begin{align}
R_{e,d}^{\mathrm{proc}}&=\sum_{v=1}^{V}\lambda_v R_v(e,d),\\
A_{e,d}^{\mathrm{proc}}&=
\frac{R_{e,d}^{\mathrm{proc}}-\mu_{G_d(q)}^{\mathrm{proc}}}
{\sigma_{G_d(q)}^{\mathrm{proc}}+\delta},
\end{align}
where $R_v(e,d)$ is a process signal with weight $\lambda_v$, and $G_d(q)$ contains units with the same role and action type. We align rollout- and decision-level advantages to tokens by
\begin{equation}
\begin{aligned}
A_{n,t}&=m_{n,t}\Bigg[A_{e(n)}^{\mathrm{out}}+
\lambda_{\mathrm{proc}}
\sum_{d\in\mathcal{Q}_{e(n)}}\\[-2pt]
&\quad\mathbb{I}[(n,t)\in\mathcal{I}_{e(n),d}]
A_{e(n),d}^{\mathrm{proc}}\Bigg].
\end{aligned}
\label{eq:token_advantage}
\end{equation}
The final task-level training set is
\begin{equation}
\begin{aligned}
\widetilde{\mathcal{D}}_{\mathcal{H}}(q)
&=\bigcup_{\tau_e\in G(q)}
\widetilde{\mathcal{D}}_{\mathcal{H}}(\tau_e),\\
\mathcal{D}_{\mathcal{H}}(q)
&=\{(S_n,\mathbf{A}_n)\mid
S_n\in\widetilde{\mathcal{D}}_{\mathcal{H}}(q)\}.
\end{aligned}
\label{eq:structured_training_set}
\end{equation}
where $\mathbf{A}_n=(A_{n,1},\ldots,A_{n,L_n^{\mathrm{resp}}})$.


For each trainable token, let
$\rho_{n,t}=\exp(\ell_{n,t}(\theta_{r(n)})-
\ell_{n,t}^{\mathrm{roll}})$ and
$\bar\rho_{n,t}=\operatorname{clip}(\rho_{n,t},1-\epsilon,1+\epsilon)$.
Let $\mathcal{I}_e^{\mathrm{train}}$ collect positions with $m_{n,t}=1$ in
$\widetilde{\mathcal D}_{\mathcal H}(\tau_e)$. The rollout-balanced clipped
objective is
\begin{equation}
\begin{aligned}
J_{\text{\M}}
&=\frac{1}{K}\sum_{e=1}^{K}
\frac{1}{\max(|\mathcal{I}_e^{\mathrm{train}}|,1)}\\[-2pt]
&\quad\sum_{(n,t)\in\mathcal{I}_e^{\mathrm{train}}}
\min\big(
\rho_{n,t}A_{n,t},\bar\rho_{n,t}A_{n,t}\big).
\end{aligned}
\label{eq:group_clipped_objective}
\end{equation}
The per-rollout normalization prevents branch-rich rollouts from dominating
the update. Partitioning the sum into Action and Args positions gives
Eq.~\eqref{eq:framework_objective}, to which CAPO applies the routed gradients
in Eq.~\eqref{eq:routed_gradients}.

\section{Experiments}
\label{sec:experiments}

We systematically evaluate \M~along the following four dimensions:
\begin{itemize}[leftmargin=*,noitemsep,topsep=2pt]
    \item \textbf{RQ1:} Does \M~consistently outperform existing
    baselines on multi-agent reasoning and collaboration benchmarks?
    \item \textbf{RQ2:} How do CAPO and its conflict-aware gradient-routing
    mechanism contribute to overall performance?
    \item \textbf{RQ3:} How does the training scope, central-only training
    versus joint training of all agents, affect system performance?
    \item \textbf{RQ4:} How do the planning and dispatch behaviors of
    \M~qualitatively differ from those of the baselines?
\end{itemize}

\subsection{Experimental Setup}

\textbf{\ding{182} Training Data.}
We construct RL queries from the training splits of 2WikiMultiHopQA~\cite{xanh2020_2wikimultihop} and HotpotQA~\cite{yang2018hotpotqa}. We remove queries that require no retrieval or only a trivial single retrieval, retaining examples that require multi-step reasoning and sub-agent coordination.
\textbf{\ding{183} Evaluation Benchmarks.} 
We report token-level F1~\cite{yacouby2020probabilistic} (\%) on seven benchmarks: 2WikiMultiHopQA~\cite{xanh2020_2wikimultihop}, HotpotQA~\cite{yang2018hotpotqa}, Bamboogle~\cite{press2022measuring}, FRAMES~\cite{krishna2025fact}, MuSiQue~\cite{trivedi2022musique}, Natural Questions~\cite{kwiatkowski2019natural}, and TriviaQA~\cite{joshi2017triviaqa}. 2WikiMultiHopQA and HotpotQA are in-domain; the remaining five are out-of-domain. This collection measures multi-hop reasoning, fact retrieval, and transfer to unseen question distributions.
\textbf{\ding{184} Backbones and agents.}
We use Qwen2.5-1.5B and Qwen2.5-3B~\cite{yang2024qwen2} as policy backbones. The harness contains a Search Agent equipped with Wikipedia and knowledge-graph retrieval tools and a Conclusion Agent equipped with an answer-synthesis tool. Unless otherwise stated, these sub-agents are frozen environment components and only the central agent is optimized. The agent configuration follows STACKPLANNER~\cite{zhang2026stackplanner}.
\textbf{\ding{185} Baselines.}
We compare four families: (i) no-retrieval methods (Base and chain-of-thought~\cite{wei2022chainofthought}); (ii) naive RAG (FS-RAG and FL-RAG); (iii) agentic RAG (ReAct~\cite{react}, IRCoT~\cite{trivedi2023interleaving}, and TC-RAG~\cite{jiang2024tcrag}); and (iv) RL-based agents (ReSearch~\cite{chen2025learning}, Search-R1~\cite{jin2025search}, AEPO~\cite{dong2025agentic}, ARPO~\cite{dong2025agentic2}, MEM1~\cite{zhou2025mem1}, and AgenticRAG-R1~\cite{Agentic_RAG_R1}).
\textbf{\ding{186} Training details.}
Each query produces $K=4$ rollouts. We set the process-reward coefficient to $\lambda_{\mathrm{proc}}=0.1$. And we train for 100 optimizer updates using four NVIDIA A100-SXM4-80GB GPUs for Qwen2.5-1.5B-Instruct and eight for Qwen2.5-3B-Instruct. Complete hyperparameters are provided in Appendix~\ref{Training_Configuration}.

\subsection{Main Result Analysis}
\label{Main_Result_Analysis}
\begin{table*}[!t]
\centering
\fontsize{9pt}{10.5pt}\selectfont
\setlength{\tabcolsep}{3pt}
\renewcommand{\arraystretch}{0.95}
\resizebox{\textwidth}{!}{
\begin{tabular}{l|l|rr|rrrrr|r}
\toprule
\rowcolor{gray!30}
\multicolumn{2}{c|}{\textbf{Method}} &
\multicolumn{2}{c|}{\textbf{In-Domain}} &
\multicolumn{5}{c|}{\textbf{Out-of-Domain}} & {} \\
\rowcolor{gray!30}
\textbf{Paradigm} & \textbf{Approach} &
\textbf{2Wiki} & \textbf{HotpotQA} & \textbf{Bamboogle} &
\textbf{FRAMES} & \textbf{MuSiQue} & \textbf{NQ} &
\textbf{TriviaQA} & \textbf{Avg.} \\
\midrule
\rowcolor{gray!10}
\multicolumn{10}{c}{\textbf{\textit{Qwen2.5-1.5B}}} \\
\midrule
\multirow{2}{*}{No RAG} & Base & 21.95 & 20.20 & 5.87 & 9.51 & 8.26 & 11.31 & 32.81 & 15.70 \\
& CoT & 16.75 & 17.90 & 19.74 & 8.43 & 7.22 & 9.82 & 25.17 & 15.00 \\
\midrule
\multirow{2}{*}{Naive RAG} & FS-RAG & 21.37 & 26.55 & 15.31 & 11.40 & 9.50 & 17.74 & 44.73 & 20.94 \\
& FL-RAG & 26.08 & 28.14 & 14.84 & 12.58 & 10.30 & 21.94 & 48.59 & 23.21 \\
\midrule
\multirow{3}{*}{Agentic RAG} & ReAct & 12.43 & 23.48 & 12.54 & 8.63 & 7.99 & 21.06 & 33.91 & 17.15 \\
& IRCoT & 21.79 & 26.68 & 17.49 & 9.32 & 8.73 & 22.93 & 44.19 & 21.59 \\
& TC-RAG & 25.23 & 22.29 & 11.63 & 10.27 & 7.90 & 14.21 & 34.05 & 17.94 \\
\midrule
\multirow{6}{*}{RL-based} & ReSearch & 25.55 & 30.45 & \underline{22.16} & 9.21 & 9.54 & 29.07 & 44.65 & 24.38 \\
& Search-R1 & 26.51 & 20.66 & 13.80 & 9.65 & 8.36 & 11.93 & 25.76 & 16.67 \\
& AEPO & 17.34 & 14.75 & 11.58 & 8.98 & 7.11 & 9.55 & 23.02 & 13.19 \\
& ARPO & 20.02 & 15.48 & 12.38 & 9.11 & 6.16 & 9.19 & 23.96 & 13.76 \\
& MEM1 & 14.43 & 14.34 & 12.03 & 8.18 & 6.08 & 18.41 & 22.06 & 13.65 \\
& AgenticRAG-R1 & \underline{29.53} & \underline{30.88} & 20.00 &
\underline{13.46} & \underline{16.48} & \underline{31.35} &
\underline{51.83} & \underline{27.65} \\
\midrule
\rowcolor[HTML]{F0F6FF}
Ours & \textbf{\M} & \textbf{41.63} & \textbf{44.44} &
\textbf{28.63} & \textbf{37.71} & \textbf{27.83} & \textbf{62.04} &
\textbf{58.22} & \textbf{42.93} \\
\midrule
\rowcolor{gray!10}
\multicolumn{10}{c}{\textbf{\textit{Qwen2.5-3B}}} \\
\midrule
\multirow{2}{*}{No RAG} & Base & 23.98 & 24.08 & 9.45 & 8.01 & 9.70 & 14.27 & 38.84 & 18.33 \\
& CoT & 18.90 & 23.82 & 20.80 & 7.16 & 10.47 & 16.53 & 39.62 & 19.61 \\
\midrule
\multirow{2}{*}{Naive RAG} & FS-RAG & 15.47 & 25.85 & 10.48 & 10.42 & 7.64 & 19.84 & 45.38 & 19.30 \\
& FL-RAG & 16.80 & 26.78 & 11.05 & 9.19 & 7.29 & 21.93 & 48.50 & 20.22 \\
\midrule
\multirow{3}{*}{Agentic RAG} & ReAct & 25.09 & 34.37 & 24.86 & 10.53 & 13.92 & 27.19 & 46.04 & 26.00 \\
& IRCoT & 15.89 & 24.50 & 25.27 & 6.79 & 12.43 & 27.86 & 49.19 & 23.13 \\
& TC-RAG & 28.47 & 21.94 & 17.59 & 7.69 & 8.99 & 20.74 & 50.46 & 22.27 \\
\midrule
\multirow{6}{*}{RL-based} & ReSearch & 27.23 & 33.96 & 15.09 & 10.00 & 9.47 & 34.61 & 53.93 & 26.33 \\
& Search-R1 & 29.90 & 37.24 & 29.90 & 10.76 & 13.53 & 34.73 & 55.08 & 30.16 \\
& AEPO & 23.01 & 28.71 & 22.09 & 12.26 & 11.70 & 26.76 & 47.78 & 24.62 \\
& ARPO & 29.55 & 36.48 & 27.32 & 13.49 & 13.38 & 33.29 & 53.66 & 29.60 \\
& MEM1 & 18.06 & 20.15 & 5.19 & 4.99 & 4.47 & 19.18 & 33.09 & 15.02 \\
& AgenticRAG-R1 & \underline{32.92} & \underline{44.00} &
\underline{31.48} & \underline{16.23} & \underline{16.48} &
\underline{37.15} & \underline{56.62} & \underline{33.55} \\
\midrule
\rowcolor[HTML]{F0F6FF}
Ours & \textbf{\M} & \textbf{45.78} & \textbf{57.35} &
\textbf{40.37} & \textbf{34.90} & \textbf{24.33} & \textbf{63.70} &
\textbf{68.07} & \textbf{47.79} \\
\bottomrule
\end{tabular}
}
\caption{Performance comparison (token-level F1, \%) on in-domain and
out-of-domain benchmarks. \textbf{Bold} denotes the best result and
\underline{underlining} denotes the strongest baseline for each backbone.}
\label{tab:main_results}
\end{table*}

\begin{table*}[!t]
\centering
\fontsize{9pt}{10.5pt}\selectfont
\setlength{\tabcolsep}{4pt}
\renewcommand{\arraystretch}{1.05}
\resizebox{\textwidth}{!}{
\begin{tabular}{l|l|rrrrrrr|r}
\toprule
\rowcolor{gray!30}
\textbf{Backbone} & \textbf{Method} & \textbf{2Wiki} & \textbf{HotpotQA} &
\textbf{Bamboogle} & \textbf{FRAMES} & \textbf{MuSiQue} & \textbf{NQ} &
\textbf{TriviaQA} & \textbf{Avg.} \\
\midrule
\rowcolor[HTML]{F0F6FF}
\cellcolor{white}\multirow{2}{*}{Qwen2.5-1.5B} & \textbf{\M} & \textbf{41.63} &
\textbf{44.44} & 28.63 & \textbf{37.71} & \textbf{27.83} &
\textbf{62.04} & \textbf{58.22} & \textbf{42.93} \\
& \textit{w/o CAPO} & 41.33 & 43.70 & \textbf{30.74} & 32.68 & 23.02 &
59.44 & 56.37 & 41.04 \\
\midrule
\rowcolor[HTML]{F0F6FF}
\cellcolor{white}\multirow{2}{*}{Qwen2.5-3B} & \textbf{\M} & 45.78 & \textbf{57.35} &
\textbf{40.37} & \textbf{34.90} & 24.33 & \textbf{63.70} &
\textbf{68.07} & \textbf{47.79} \\
& \textit{w/o CAPO} & \textbf{47.56} & 53.70 & 38.89 & 32.67 &
\textbf{25.02} & 60.37 & 65.31 & 46.22 \\
\bottomrule
\end{tabular}
}
\caption{Ablation of CAPO measured by token-level F1 (\%). Blue rows denote
the complete method, and \textbf{bold} indicates the better result within each
backbone.}
\label{tab:capo_ablation}
\end{table*}

\begin{table*}[!t]
\centering
\fontsize{9pt}{10.5pt}\selectfont
\setlength{\tabcolsep}{4pt}
\renewcommand{\arraystretch}{1.05}
\resizebox{\textwidth}{!}{
\begin{tabular}{l|l|rrrrrrr|r}
\toprule
\rowcolor{gray!30}
\textbf{Metric} & \textbf{Training Scope} & \textbf{2Wiki} &
\textbf{HotpotQA} & \textbf{Bamboogle} & \textbf{FRAMES} &
\textbf{MuSiQue} & \textbf{NQ} & \textbf{TriviaQA} & \textbf{Avg.} \\
\midrule
\rowcolor[HTML]{F0F6FF}
\cellcolor{white}\multirow{2}{*}{Accuracy} & \textbf{Central only} & 42.22 & \textbf{55.56} &
\textbf{26.67} & \textbf{28.89} & \textbf{17.78} & \textbf{61.67} &
67.34 & \textbf{42.88} \\
& All agents & \textbf{45.34} & 53.33 & 22.22 & 22.56 & 15.56 & 55.56 &
\textbf{68.89} & 40.49 \\
\midrule
\rowcolor[HTML]{F0F6FF}
\cellcolor{white}\multirow{2}{*}{F1} & \textbf{Central only} & \textbf{45.78} &
\textbf{57.35} & \textbf{40.37} & \textbf{34.90} & 24.33 &
\textbf{63.70} & \textbf{68.07} & \textbf{47.79} \\
& All agents & 43.24 & 52.59 & 38.15 & 33.54 & \textbf{24.86} & 58.78 &
66.38 & 45.36 \\
\bottomrule
\end{tabular}
}
\caption{Effect of training scope with Qwen2.5-3B. Blue rows denote
central-only training, and \textbf{bold} indicates the better result under each
metric.}
\label{tab:training_scope}
\end{table*}

\begin{figure*}[t]
    \centering
    \includegraphics[width=0.8\textwidth]{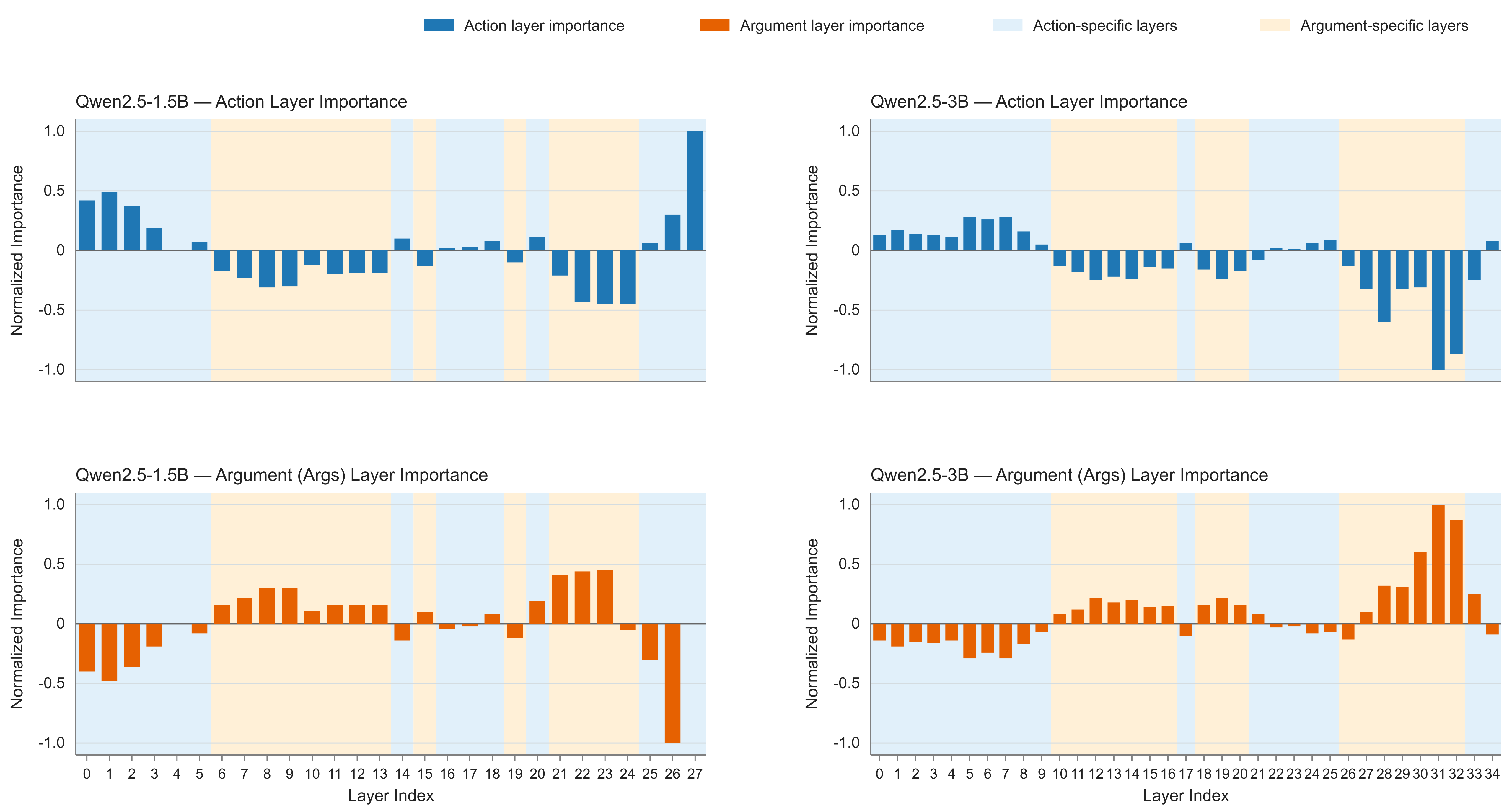}
    \caption{Layer-wise Action/Args functional partitions derived from forward
    activations. Blue and orange denote Action and Args importance and their
    selected layers, respectively. Left: Qwen2.5-1.5B (28 layers). Right:
    Qwen2.5-3B (36 layers).}
    \label{fig:functional_partition}
    \vspace{-0.5cm}
\end{figure*}

Table~\ref{tab:main_results} compares \M~with the baselines on the multi-hop QA and agentic retrieval benchmarks. With Qwen2.5-1.5B, \M~achieves an average F1 of 42.93, outperforming AgenticRAG-R1, the strongest baseline overall, by 15.28 points and the strong RL-based ReSearch baseline by 18.55 points. Specifically, \M~obtains F1 scores of 41.63, 44.44, 28.63, 37.71, 27.83, 62.04, and 58.22 on 2Wiki, HotpotQA, Bamboogle, FRAMES, MuSiQue, NQ, and TriviaQA, respectively, achieving the best result on all seven benchmarks. With Qwen2.5-3B, \M~reaches an average F1 of 47.79, exceeding the strongest baseline, AgenticRAG-R1, by 14.24 points. It also outperforms Search-R1 and ARPO by 17.63 and 18.19 points, respectively. On 2Wiki, HotpotQA, Bamboogle, FRAMES, MuSiQue, NQ, and TriviaQA, \M~achieves F1 scores of 45.78, 57.35, 40.37, 34.90, 24.33, 63.70, and 68.07, respectively, again ranking first on every benchmark. These results show that \M~consistently improves the reasoning, dispatch, and coordination capabilities of the central agent across both model scales, providing an affirmative answer to \textbf{RQ1}.

\subsection{Component Analysis}
To answer \textbf{RQ2--RQ3}, we compare CAPO with standard GRPO and central-only training with joint multi-agent training.

\textbf{\ding{182} Effectiveness of CAPO.}
As shown in Table~\ref{tab:capo_ablation}, \M~achieves an average F1 of 42.93 with Qwen2.5-1.5B, exceeding \textit{w/o CAPO} (41.04) by 1.89 points. With Qwen2.5-3B, it improves the average F1 from 46.22 to 47.79, a gain of 1.57 points. CAPO produces clear improvements on FRAMES, NQ, and TriviaQA at both model scales. Its effect nevertheless varies across datasets: the complete method underperforms \textit{w/o CAPO} on Bamboogle with Qwen2.5-1.5B and on 2Wiki and MuSiQue with Qwen2.5-3B. One possible explanation is that the training data are primarily drawn from 2WikiMultiHopQA, whereas the evaluation benchmarks differ in question decomposition, reasoning-path length, and the difficulty of args generation. Consequently, the effectiveness of decoupled optimization may vary across data distributions. Overall, the improvement in average F1 demonstrates the contribution of CAPO and its conflict-aware gradient-routing mechanism, answering \textbf{RQ2}.

\textbf{\ding{183} Functional Partition Analysis.}
To further examine the behavior of CAPO, Figure~\ref{fig:functional_partition} visualizes the normalized layer-wise importance used to construct the Action and Args functional partitions. For both Qwen2.5-1.5B and Qwen2.5-3B, Action Tokens and Args Tokens exhibit different importance distributions across layers, resulting in distinct selected partitions. This observation supports the use of activation-based functional partitioning to separate the two heterogeneous decision types.


\textbf{\ding{184} Effect of Training Scope.}
Table~\ref{tab:training_scope} compares \textit{Central only}, which trains only the central agent, with \textit{All Agents}, which jointly trains the central agent and sub-agents. Both configurations use decoupled policy optimization with Qwen2.5-3B. In terms of accuracy, their average scores are 42.88 and 40.49, respectively, giving central-only training a 2.38-point advantage. In terms of F1, central-only and all-agent training achieve 47.79 and 45.36, respectively, a difference of 2.42 points. Although joint training is better on a small number of individual benchmarks, it does not improve overall performance. These results suggest that, in the evaluated central-coordination tasks, optimizing only the central agent better preserves the consistency of the global planning objective and reduces interference among policy updates from different agents, answering \textbf{RQ3}.

\section{Conclusion}
 We introduced  \M, a reinforcement learning framework for central-agent policies in structured multi-agent harnesses. Its interface-level black-box mechanism reconstructs multi-round, multi-session, and branching interactions as aligned training samples, then combines rollout-level outcome rewards with decision-level process rewards. CAPO complements this representation by using forward activations to identify action- and args-related parameter subsets and routing their policy gradients accordingly.  Across seven multi-hop QA and agentic retrieval benchmarks, \M~achieves average F1 scores of 42.93 with Qwen2.5-1.5B and 47.79 with Qwen2.5-3B, outperforming the strongest non-ours row in the reported comparison by 15.28 and 14.24 points, respectively. Removing CAPO reduces average F1 by 1.89 and 1.57 points, and central-only training outperforms joint training in the evaluated 3B harness. Future work should test larger policies, more diverse harness topologies, dynamic functional partitions, multimodal actions, and long-term memory, while measuring the robustness and computational cost of partition estimation.

\section*{Ethical considerations}
To evaluate the proposed method, we conducted experiments exclusively on publicly available benchmark datasets, including 2WikiMultiHopQA, HotpotQA, Bamboogle, FRAMES, MuSiQue, Natural Questions, and TriviaQA, in accordance with their respective licenses and usage policies. We did not intentionally collect or use any personally identifiable information, and no human or animal subjects were involved in this research.

\bibliography{custom}

@String{Computing = "Computing" }

@String{Computer = "{IEEE} Computer" }

@ArtifactSoftware{R,
    title = {R: A Language and Environment for Statistical Computing},
    author = {{R Core Team}},
    organization = {R Foundation for Statistical Computing},
    address = {Vienna, Austria},
    year = {2019},
    url = {https://www.R-project.org/},
}

@misc{f1,
      title={Retrieving and Reading: A Comprehensive Survey on Open-domain Question Answering}, 
      author={Fengbin Zhu and Wenqiang Lei and Chao Wang and Jianming Zheng and Soujanya Poria and Tat-Seng Chua},
      year={2021},
      eprint={2101.00774},
      archivePrefix={arXiv},
      primaryClass={cs.AI}
}

@article{dai2021knowledge,
  title={Knowledge neurons in pretrained transformers},
  author={Dai, Damai and Dong, Li and Hao, Yaru and Sui, Zhifang and Chang, Baobao and Wei, Furu},
  journal={arXiv preprint arXiv:2104.08696},
  year={2021}
}

@article{yao2022react,
  title={React: Synergizing reasoning and acting in language models},
  author={Yao, Shunyu and Zhao, Jeffrey and Yu, Dian and Du, Nan and Shafran, Izhak and Narasimhan, Karthik and Cao, Yuan},
  journal={arXiv preprint arXiv:2210.03629},
  year={2022}
}

@article{todd2023function,
  title={Function vectors in large language models},
  author={Todd, Eric and Li, Millicent L and Sharma, Arnab Sen and Mueller, Aaron and Wallace, Byron C and Bau, David},
  journal={arXiv preprint arXiv:2310.15213},
  year={2023}
}

@inproceedings{lape,
    title = "Language-Specific Neurons: The Key to Multilingual Capabilities in Large Language Models",
    author = "Tang, Tianyi  and
      Luo, Wenyang  and
      Huang, Haoyang  and
      Zhang, Dongdong  and
      Wang, Xiaolei  and
      Zhao, Xin  and
      Wei, Furu  and
      Wen, Ji-Rong",
    editor = "Ku, Lun-Wei  and
      Martins, Andre  and
      Srikumar, Vivek",
    booktitle = "Proceedings of the 62nd Annual Meeting of the Association for Computational Linguistics (Volume 1: Long Papers)",
    month = aug,
    year = "2024",
    address = "Bangkok, Thailand",
    publisher = "Association for Computational Linguistics",
    url = "https://aclanthology.org/2024.acl-long.309/",
    doi = "10.18653/v1/2024.acl-long.309",
    pages = "5701--5715"
}

@inproceedings{yacouby2020probabilistic,
  title={Probabilistic extension of precision, recall, and f1 score for more thorough evaluation of classification models},
  author={Yacouby, Reda and Axman, Dustin},
  booktitle={Proceedings of the first workshop on evaluation and comparison of NLP systems},
  pages={79--91},
  year={2020}
}

@inproceedings{jiang2025model,
  title={MODEL SHAPLEY: Find Your Ideal Parameter Player via One Gradient Backpropagation},
  author={Chu, Xu and Jiang, Xinke and Qiu, Rihong and Gao, Jiaran and Zhao, Junfeng},
  booktitle={NeurIPS 2025}
}

@article{hu2021lora,
  author = {Hu, E. J. and Shen, Y. and Wallis, P. and Allen-Zhu, Z. and Li, Y. and Wang, S. and Wang, L. and Chen, W.},
  title = {{LoRA}: Low-Rank Adaptation of Large Language Models},
  year = {2021},
  eprint = {2106.09685},
  eprinttype = {arxiv},
  url = {https://arxiv.org/abs/2106.09685}
}

@misc{qian2025toolrlrewardtoollearning,
      title={ToolRL: Reward is All Tool Learning Needs}, 
      author={Cheng Qian and Emre Can Acikgoz and Qi He and Hongru Wang and Xiusi Chen and Dilek Hakkani-Tür and Gokhan Tur and Heng Ji},
      year={2025},
      eprint={2504.13958},
      archivePrefix={arXiv},
      primaryClass={cs.LG},
      url={https://arxiv.org/abs/2504.13958}, 
}

@misc{wang2025ragenunderstandingselfevolutionllm,
      title={RAGEN: Understanding Self-Evolution in LLM Agents via Multi-Turn Reinforcement Learning}, 
      author={Zihan Wang and Kangrui Wang and Qineng Wang and Pingyue Zhang and Linjie Li and Zhengyuan Yang and Xing Jin and Kefan Yu and Minh Nhat Nguyen and Licheng Liu and Eli Gottlieb and Yiping Lu and Kyunghyun Cho and Jiajun Wu and Li Fei-Fei and Lijuan Wang and Yejin Choi and Manling Li},
      year={2025},
      eprint={2504.20073},
      archivePrefix={arXiv},
      primaryClass={cs.LG},
      url={https://arxiv.org/abs/2504.20073}, 
}

@article{jin2025search,
  title = {{Search-R1}: Training {LLMs} to Reason and Leverage Search Engines with Reinforcement Learning},
  author = {Jin, Bowen and Zeng, Hansi and Yue, Zhenrui and Yoon, Jinsung and Arik, Sercan and Wang, Dong and Zamani, Hamed and Han, Jiawei},
  journal = {arXiv preprint arXiv:2503.09516},
  year = {2025}
}

@inproceedings{yang2018hotpotqa,
  title={HotpotQA: A Dataset for Diverse, Explainable Multi-hop Question Answering},
  author={Yang, Zhilin and Qi, Peng and Zhang, Saizheng and others},
  booktitle={EMNLP},
  year={2018}
}

@article{trivedi2022musique,
  title={MuSiQue: Multihop Questions via Single-hop Question Composition},
  author={Trivedi, Harsh and Balasubramanian, Niranjan and Khot, Tushar and Sabharwal, Ashish},
  journal={Transactions of the Association for Computational Linguistics},
  volume={10},
  pages={539--554},
  year={2022},
  publisher={MIT Press One Broadway, 12th Floor, Cambridge, Massachusetts 02142, USA~…}
}

@inproceedings{krishna2025fact,
  title={Fact, fetch, and reason: A unified evaluation of retrieval-augmented generation},
  author={Krishna, Satyapriya and Krishna, Kalpesh and Mohananey, Anhad and Schwarcz, Steven and Stambler, Adam and Upadhyay, Shyam and Faruqui, Manaal},
  booktitle={Proceedings of the 2025 Conference of the Nations of the Americas Chapter of the Association for Computational Linguistics: Human Language Technologies (Volume 1: Long Papers)},
  pages={4745--4759},
  year={2025}
}

@article{kwiatkowski2019natural,
  title={Natural Questions: A Benchmark for Question Answering Research},
  author={Kwiatkowski, Tom and others},
  journal={TACL},
  year={2019}
}

@inproceedings{joshi2017triviaqa,
  title={TriviaQA: A Large Scale Distantly Supervised Challenge Dataset for Reading Comprehension},
  author={Joshi, Mandar and others},
  booktitle={ACL},
  year={2017}
}

@article{chen2025learning,
  title={Learning to reason with search for llms via reinforcement learning},
  author={Chen, Mingyang and Sun, Linzhuang and Li, Tianpeng and Sun, Haoze and Zhou, Yijie and Zhu, Chenzheng and Wang, Haofen and Pan, Jeff Z and Zhang, Wen and Chen, Huajun and others},
  journal={arXiv preprint arXiv:2503.19470},
  year={2025}
}

@inproceedings{trivedi2023interleaving,
  title={Interleaving Retrieval with Chain-of-Thought Reasoning for Knowledge-Intensive Multi-Step Questions},
  author={Trivedi, Harsh and others},
  booktitle={ACL},
  year={2023}
}

@article{zhou2025mem1,
  title={{MEM1}: Learning to Synergize Memory and Reasoning for Efficient Long-Horizon Agents},
  author={Zhou, Zijian and Qu, Ao and Wu, Zhaoxuan and Kim, Sunghwan and Prakash, Alok and Rus, Daniela and Zhao, Jinhua and Low, Bryan Kian Hsiang and Liang, Paul Pu},
  journal={arXiv preprint arXiv:2506.15841},
  year={2025}
}

@article{react,
  title={ReAct: Synergizing Reasoning and Acting in Language Models},
  author={Yao, Shunyu and Zhao, Jeffrey and Yu, Dian and Du, Nan and Shafran, Izhak and Narasimhan, Karthik and Cao, Yuan},
  journal={arXiv preprint arXiv:2210.03629},
  year={2022}
}

@online{openaio1,
  author = {OpenAI},
  title = {OpenAI-o1},
  year = 2025,
  url = {https://openai.com/o1/},
  note = {Accessed: 2025-05-16}
}

@article{yang2024qwen2,
  title={Qwen2. 5 technical report},
  author={Yang, An and Yang, Baosong and Zhang, Beichen and Hui, Binyuan and Zheng, Bo and Yu, Bowen and Li, Chengyuan and Liu, Dayiheng and Huang, Fei and Wei, Haoran and others},
  journal={arXiv preprint arXiv:2412.15115},
  year={2024}
}

@article{jiang2024tcrag,
  title={TC-RAG: Turing-Complete RAG's Case study on Medical LLM Systems},
  author={Jiang, Xinke and Fang, Yue and Qiu, Rihong and Zhang, Haoyu and Xu, Yongxin and Chen, Hao and Zhang, Wentao and Zhang, Ruizhe and Fang, Yuchen and Chu, Xu and others},
  journal={arXiv preprint arXiv:2408.09199},
  year={2024}
}

@article{guo2025deepseek,
  title={Deepseek-r1: Incentivizing reasoning capability in llms via reinforcement learning},
  author={Guo, Daya and Yang, Dejian and Zhang, Haowei and Song, Junxiao and Zhang, Ruoyu and Xu, Runxin and Zhu, Qihao and Ma, Shirong and Wang, Peiyi and Bi, Xiao and others},
  journal={arXiv preprint arXiv:2501.12948},
  year={2025}
}

@article{dong2025agentic,
  title={Agentic entropy-balanced policy optimization},
  author={Dong, Guanting and Bao, Licheng and Wang, Zhongyuan and Zhao, Kangzhi and Li, Xiaoxi and Jin, Jiajie and Yang, Jinghan and Mao, Hangyu and Zhang, Fuzheng and Gai, Kun and others},
  journal={arXiv preprint arXiv:2510.14545},
  year={2025}
}

@article{dong2025agentic2,
  title={Agentic reinforced policy optimization},
  author={Dong, Guanting and Mao, Hangyu and Ma, Kai and Bao, Licheng and Chen, Yifei and Wang, Zhongyuan and Chen, Zhongxia and Du, Jiazhen and Wang, Huiyang and Zhang, Fuzheng and others},
  journal={arXiv preprint arXiv:2507.19849},
  year={2025}
}

@article{wei2022chainofthought,
  title={Chain-of-thought prompting elicits reasoning in large language models},
  author={Wei, Jason and Wang, Xuezhi and Schuurmans, Dale and Bosma, Maarten and Xia, Fei and Chi, Ed and Le, Quoc V and Zhou, Denny and others},
  journal={Advances in neural information processing systems},
  volume={35},
  pages={24824--24837},
  year={2022}
}

@article{shao2024deepseekmath,
  title={Deepseekmath: Pushing the limits of mathematical reasoning in open language models},
  author={Shao, Zhihong and Wang, Peiyi and Zhu, Qihao and Xu, Runxin and Song, Junxiao and Bi, Xiao and Zhang, Haowei and Zhang, Mingchuan and Li, YK and Wu, Y and others},
  journal={arXiv preprint arXiv:2402.03300},
  year={2024}
}

@article{press2022measuring,
  title={Measuring and narrowing the compositionality gap in language models},
  author={Press, Ofir and Zhang, Muru and Min, Sewon and Schmidt, Ludwig and Smith, Noah A and Lewis, Mike},
  journal={arXiv preprint arXiv:2210.03350},
  year={2022}
}

@inproceedings{xanh2020_2wikimultihop,
    title = "Constructing A Multi-hop {QA} Dataset for Comprehensive Evaluation of Reasoning Steps",
    author = "Ho, Xanh  and
      Duong Nguyen, Anh-Khoa  and
      Sugawara, Saku  and
      Aizawa, Akiko",
    booktitle = "Proceedings of the 28th International Conference on Computational Linguistics",
    month = dec,
    year = "2020",
    address = "Barcelona, Spain (Online)",
    publisher = "International Committee on Computational Linguistics",
    url = "https://www.aclweb.org/anthology/2020.coling-main.580",
    pages = "6609--6625",
}

@article{DBLP:journals/corr/abs-2406-13381,
  author       = {Xinming Hou and
                  Mingming Yang and
                  Wenxiang Jiao and
                  Xing Wang and
                  Zhaopeng Tu and
                  Wayne Xin Zhao},
  title        = {CoAct: {A} Global-Local Hierarchy for Autonomous Agent Collaboration},
  journal      = {CoRR},
  volume       = {abs/2406.13381},
  year         = {2024},
  url          = {https://doi.org/10.48550/arXiv.2406.13381},
  doi          = {10.48550/ARXIV.2406.13381},
  eprinttype   = {arXiv},
  eprint       = {2406.13381},
  bibsource    = {dblp computer science bibliography, https://dblp.org}
}

@inproceedings{DBLP:conf/acl/YueZLWWCQ25,
  author       = {Yanwei Yue and
                  Guibin Zhang and
                  Boyang Liu and
                  Guancheng Wan and
                  Kun Wang and
                  Dawei Cheng and
                  Yiyan Qi},
  editor       = {Wanxiang Che and
                  Joyce Nabende and
                  Ekaterina Shutova and
                  Mohammad Taher Pilehvar},
  title        = {MasRouter: Learning to Route LLMs for Multi-Agent Systems},
  booktitle    = {Proceedings of the 63rd Annual Meeting of the Association for Computational
                  Linguistics (Volume 1: Long Papers), {ACL} 2025, Vienna, Austria,
                  July 27 - August 1, 2025},
  pages        = {15549--15572},
  publisher    = {Association for Computational Linguistics},
  year         = {2025},
  url          = {https://doi.org/10.18653/v1/2025.acl-long.757},
  doi          = {10.18653/V1/2025.ACL-LONG.757},
  bibsource    = {dblp computer science bibliography, https://dblp.org}
}

@article{zhang2026stackplanner,
  title={StackPlanner: A Centralized Hierarchical Multi-Agent System with Task-Experience Memory Management},
  author={Zhang, Ruizhe and Jiang, Xinke and Yang, Zhibang and Zhang, Zhixin and Gao, Jiaran and Xiao, Yuzhen and Lai, Hongbin and Chu, Xu and Zhao, Junfeng and Wang, Yasha},
  journal={arXiv preprint arXiv:2601.05890},
  year={2026}
}

@article{DBLP:journals/corr/abs-2308-08155,
  author       = {Qingyun Wu and
                  Gagan Bansal and
                  Jieyu Zhang and
                  Yiran Wu and
                  Shaokun Zhang and
                  Erkang Zhu and
                  Beibin Li and
                  Li Jiang and
                  Xiaoyun Zhang and
                  Chi Wang},
  title        = {AutoGen: Enabling Next-Gen {LLM} Applications via Multi-Agent Conversation
                  Framework},
  journal      = {CoRR},
  volume       = {abs/2308.08155},
  year         = {2023},
  url          = {https://doi.org/10.48550/arXiv.2308.08155},
  doi          = {10.48550/ARXIV.2308.08155},
  eprinttype   = {arXiv},
  eprint       = {2308.08155},
  bibsource    = {dblp computer science bibliography, https://dblp.org}
}

@inproceedings{DBLP:conf/iclr/HongZCZCWZWYLZR24,
  author       = {Sirui Hong and
                  Mingchen Zhuge and
                  Jonathan Chen and
                  Xiawu Zheng and
                  Yuheng Cheng and
                  Jinlin Wang and
                  Ceyao Zhang and
                  Zili Wang and
                  Steven Ka Shing Yau and
                  Zijuan Lin and
                  Liyang Zhou and
                  Chenyu Ran and
                  Lingfeng Xiao and
                  Chenglin Wu and
                  J{\"{u}}rgen Schmidhuber},
  title        = {MetaGPT: Meta Programming for {A} Multi-Agent Collaborative Framework},
  booktitle    = {The Twelfth International Conference on Learning Representations,
                  {ICLR} 2024, Vienna, Austria, May 7-11, 2024},
  publisher    = {OpenReview.net},
  year         = {2024},
  url          = {https://openreview.net/forum?id=VtmBAGCN7o},
  bibsource    = {dblp computer science bibliography, https://dblp.org}
}

@inproceedings{DBLP:conf/acl/QianLLCDL0CSCXL24,
  author       = {Chen Qian and
                  Wei Liu and
                  Hongzhang Liu and
                  Nuo Chen and
                  Yufan Dang and
                  Jiahao Li and
                  Cheng Yang and
                  Weize Chen and
                  Yusheng Su and
                  Xin Cong and
                  Juyuan Xu and
                  Dahai Li and
                  Zhiyuan Liu and
                  Maosong Sun},
  editor       = {Lun{-}Wei Ku and
                  Andre Martins and
                  Vivek Srikumar},
  title        = {ChatDev: Communicative Agents for Software Development},
  booktitle    = {Proceedings of the 62nd Annual Meeting of the Association for Computational
                  Linguistics (Volume 1: Long Papers), {ACL} 2024, Bangkok, Thailand,
                  August 11-16, 2024},
  pages        = {15174--15186},
  publisher    = {Association for Computational Linguistics},
  year         = {2024},
  url          = {https://doi.org/10.18653/v1/2024.acl-long.810},
  doi          = {10.18653/V1/2024.ACL-LONG.810},
  bibsource    = {dblp computer science bibliography, https://dblp.org}
}

@inproceedings{DBLP:conf/ijcai/GuoCWCPCW024,
  author       = {Taicheng Guo and
                  Xiuying Chen and
                  Yaqi Wang and
                  Ruidi Chang and
                  Shichao Pei and
                  Nitesh V. Chawla and
                  Olaf Wiest and
                  Xiangliang Zhang},
  title        = {Large Language Model Based Multi-agents: {A} Survey of Progress and
                  Challenges},
  booktitle    = {Proceedings of the Thirty-Third International Joint Conference on
                  Artificial Intelligence, {IJCAI} 2024, Jeju, South Korea, August 3-9,
                  2024},
  pages        = {8048--8057},
  publisher    = {ijcai.org},
  year         = {2024},
  url          = {https://www.ijcai.org/proceedings/2024/890},
  bibsource    = {dblp computer science bibliography, https://dblp.org}
}

@misc{chen2025surveyllmbasedmultiagentsystem,
      title={A Survey on LLM-based Multi-Agent System: Recent Advances and New Frontiers in Application}, 
      author={Shuaihang Chen and Yuanxing Liu and Wei Han and Weinan Zhang and Ting Liu},
      year={2025},
      eprint={2412.17481},
      archivePrefix={arXiv},
      primaryClass={cs.CL},
      url={https://arxiv.org/abs/2412.17481}, 
}

@inproceedings{DBLP:conf/icml/Du00TM24,
  author       = {Yilun Du and
                  Shuang Li and
                  Antonio Torralba and
                  Joshua B. Tenenbaum and
                  Igor Mordatch},
  editor       = {Ruslan Salakhutdinov and
                  Zico Kolter and
                  Katherine A. Heller and
                  Adrian Weller and
                  Nuria Oliver and
                  Jonathan Scarlett and
                  Felix Berkenkamp},
  title        = {Improving Factuality and Reasoning in Language Models through Multiagent
                  Debate},
  booktitle    = {Forty-first International Conference on Machine Learning, {ICML} 2024,
                  Vienna, Austria, July 21-27, 2024},
  series       = {Proceedings of Machine Learning Research},
  volume       = {235},
  pages        = {11733--11763},
  publisher    = {{PMLR} / OpenReview.net},
  year         = {2024},
  url          = {https://proceedings.mlr.press/v235/du24e.html},
  bibsource    = {dblp computer science bibliography, https://dblp.org}
}

@inproceedings{DBLP:conf/nips/YangCSSQRZ25,
  author       = {Yingxuan Yang and
                  Huacan Chai and
                  Shuai Shao and
                  Yuanyi Song and
                  Siyuan Qi and
                  Renting Rui and
                  Weinan Zhang},
  editor       = {Danielle Belgrave and
                  Cheng Zhang and
                  Laura N. Montoya and
                  Hsuan{-}Tien Lin and
                  Razvan Pascanu and
                  Piotr Koniusz and
                  Marzyeh Ghassemi and
                  Nancy Chen and
                  Iv{\'{a}}n Vladimir Meza Ru{\'{\i}}z and
                  Arturo Loaiza{-}Bonilla},
  title        = {AgentNet: Decentralized Evolutionary Coordination for LLM-based Multi-Agent
                  Systems},
  booktitle    = {Advances in Neural Information Processing Systems 38: Annual Conference
                  on Neural Information Processing Systems 2025, NeurIPS 2025, San Diego,
                  CA, USA, December 2-7, 2025 / Mexico City, Mexico, November 30 - December
                  5, 2025},
  year         = {2025},
  url          = {http://papers.nips.cc/paper\_files/paper/2025/hash/9a379c1b05793d1c42dc832269834515-Abstract-Conference.html},
  bibsource    = {dblp computer science bibliography, https://dblp.org}
}

@inproceedings{DBLP:conf/iclr/QianXW0ZXDDC00025,
  author       = {Chen Qian and
                  Zihao Xie and
                  Yifei Wang and
                  Wei Liu and
                  Kunlun Zhu and
                  Hanchen Xia and
                  Yufan Dang and
                  Zhuoyun Du and
                  Weize Chen and
                  Cheng Yang and
                  Zhiyuan Liu and
                  Maosong Sun},
  title        = {Scaling Large Language Model-based Multi-Agent Collaboration},
  booktitle    = {The Thirteenth International Conference on Learning Representations,
                  {ICLR} 2025, Singapore, April 24-28, 2025},
  publisher    = {OpenReview.net},
  year         = {2025},
  url          = {https://openreview.net/forum?id=K3n5jPkrU6},
  bibsource    = {dblp computer science bibliography, https://dblp.org}
}

@inproceedings{DBLP:conf/iclr/ZhangXYTCCZCHWZ25,
  author       = {Jiayi Zhang and
                  Jinyu Xiang and
                  Zhaoyang Yu and
                  Fengwei Teng and
                  Xionghui Chen and
                  Jiaqi Chen and
                  Mingchen Zhuge and
                  Xin Cheng and
                  Sirui Hong and
                  Jinlin Wang and
                  Bingnan Zheng and
                  Bang Liu and
                  Yuyu Luo and
                  Chenglin Wu},
  title        = {AFlow: Automating Agentic Workflow Generation},
  booktitle    = {The Thirteenth International Conference on Learning Representations,
                  {ICLR} 2025, Singapore, April 24-28, 2025},
  publisher    = {OpenReview.net},
  year         = {2025},
  url          = {https://openreview.net/forum?id=z5uVAKwmjf},
  bibsource    = {dblp computer science bibliography, https://dblp.org}
}

@inproceedings{DBLP:conf/nips/HuZFNYXSJLZWYGL25,
  author       = {Mengkang Hu and
                  Yuhang Zhou and
                  Wendong Fan and
                  Yuzhou Nie and
                  Ziyu Ye and
                  Bowei Xia and
                  Tao Sun and
                  Zhaoxuan Jin and
                  Yingru Li and
                  Zeyu Zhang and
                  Yifeng Wang and
                  Qianshuo Ye and
                  Bernard Ghanem and
                  Ping Luo and
                  Guohao Li},
  editor       = {Danielle Belgrave and
                  Cheng Zhang and
                  Laura N. Montoya and
                  Hsuan{-}Tien Lin and
                  Razvan Pascanu and
                  Piotr Koniusz and
                  Marzyeh Ghassemi and
                  Nancy Chen and
                  Iv{\'{a}}n Vladimir Meza Ru{\'{\i}}z and
                  Arturo Loaiza{-}Bonilla},
  title        = {{OWL:} Optimized Workforce Learning for General Multi-Agent Assistance
                  in Real-World Task Automation},
  booktitle    = {Advances in Neural Information Processing Systems 38: Annual Conference
                  on Neural Information Processing Systems 2025, NeurIPS 2025, San Diego,
                  CA, USA, December 2-7, 2025 / Mexico City, Mexico, November 30 - December
                  5, 2025},
  year         = {2025},
  url          = {http://papers.nips.cc/paper\_files/paper/2025/hash/48dcc43a534c5b582f9d0fdb778e9b84-Abstract-Conference.html},
  bibsource    = {dblp computer science bibliography, https://dblp.org}
}

@inproceedings{DBLP:conf/nips/YuZZYZYDFLLLLLL25,
  author       = {Qiying Yu and
                  Zheng Zhang and
                  Ruofei Zhu and
                  Yufeng Yuan and
                  Xiaochen Zuo and
                  Yu Yue and
                  Weinan Dai and
                  Tiantian Fan and
                  Gaohong Liu and
                  Juncai Liu and
                  Lingjun Liu and
                  Xin Liu and
                  Haibin Lin and
                  Zhiqi Lin and
                  Bole Ma and
                  Guangming Sheng and
                  Yuxuan Tong and
                  Chi Zhang and
                  Mofan Zhang and
                  Ru Zhang and
                  Wang Zhang and
                  Hang Zhu and
                  Jinhua Zhu and
                  Jiaze Chen and
                  Jiangjie Chen and
                  Chengyi Wang and
                  Hongli Yu and
                  Yuxuan Song and
                  Xiangpeng Wei and
                  Hao Zhou and
                  Jingjing Liu and
                  Wei{-}Ying Ma and
                  Ya{-}Qin Zhang and
                  Lin Yan and
                  Yonghui Wu and
                  Mingxuan Wang},
  editor       = {Danielle Belgrave and
                  Cheng Zhang and
                  Laura N. Montoya and
                  Hsuan{-}Tien Lin and
                  Razvan Pascanu and
                  Piotr Koniusz and
                  Marzyeh Ghassemi and
                  Nancy Chen and
                  Iv{\'{a}}n Vladimir Meza Ru{\'{\i}}z and
                  Arturo Loaiza{-}Bonilla},
  title        = {{DAPO:} An Open-Source {LLM} Reinforcement Learning System at Scale},
  booktitle    = {Advances in Neural Information Processing Systems 38: Annual Conference
                  on Neural Information Processing Systems 2025, NeurIPS 2025, San Diego,
                  CA, USA, December 2-7, 2025 / Mexico City, Mexico, November 30 - December
                  5, 2025},
  year         = {2025},
  url          = {http://papers.nips.cc/paper\_files/paper/2025/hash/a4277440d50f1f15d2cb4c14f7e0c0d2-Abstract-Conference.html},
  bibsource    = {dblp computer science bibliography, https://dblp.org}
}

@inproceedings{meng2022rome,
  author       = {Kevin Meng and
                  David Bau and
                  Alex Andonian and
                  Yonatan Belinkov},
  editor       = {Sanmi Koyejo and
                  S. Mohamed and
                  A. Agarwal and
                  Danielle Belgrave and
                  K. Cho and
                  A. Oh},
  title        = {Locating and Editing Factual Associations in {GPT}},
  booktitle    = {Advances in Neural Information Processing Systems 35: Annual Conference
                  on Neural Information Processing Systems 2022, NeurIPS 2022, New Orleans,
                  LA, USA, November 28 - December 9, 2022},
  year         = {2022},
  url          = {http://papers.nips.cc/paper\_files/paper/2022/hash/6f1d43d5a82a37e89b0665b33bf3a182-Abstract-Conference.html},
  bibsource    = {dblp computer science bibliography, https://dblp.org}
}

@inproceedings{meng2023memit,
  author       = {Kevin Meng and
                  Arnab Sen Sharma and
                  Alex J. Andonian and
                  Yonatan Belinkov and
                  David Bau},
  title        = {Mass-Editing Memory in a Transformer},
  booktitle    = {The Eleventh International Conference on Learning Representations,
                  {ICLR} 2023, Kigali, Rwanda, May 1-5, 2023},
  publisher    = {OpenReview.net},
  year         = {2023},
  url          = {https://openreview.net/forum?id=MkbcAHIYgyS},
  bibsource    = {dblp computer science bibliography, https://dblp.org}
}

@inproceedings{guo2021diffpruning,
  author       = {Demi Guo and
                  Alexander M. Rush and
                  Yoon Kim},
  editor       = {Chengqing Zong and
                  Fei Xia and
                  Wenjie Li and
                  Roberto Navigli},
  title        = {Parameter-Efficient Transfer Learning with Diff Pruning},
  booktitle    = {Proceedings of the 59th Annual Meeting of the Association for Computational
                  Linguistics and the 11th International Joint Conference on Natural
                  Language Processing, {ACL/IJCNLP} 2021, (Volume 1: Long Papers), Virtual
                  Event, August 1-6, 2021},
  pages        = {4884--4896},
  publisher    = {Association for Computational Linguistics},
  year         = {2021},
  url          = {https://doi.org/10.18653/v1/2021.acl-long.378},
  doi          = {10.18653/V1/2021.ACL-LONG.378},
  bibsource    = {dblp computer science bibliography, https://dblp.org}
}

@inproceedings{cui2025freemad,
  author       = {Yu Cui and
                  Hang Fu and
                  Haibin Zhang and
                  Licheng Wang and
                  Cong Zuo},
  editor       = {Maria Liakata and
                  Viviane P. Moreira and
                  Jiajun Zhang and
                  David Jurgens},
  title        = {Free-MAD: Consensus-Free Multi-Agent Debate},
  booktitle    = {Findings of the Association for Computational Linguistics, {ACL} 2026,
                  San Diego, California, United States, July 2-7, 2026},
  pages        = {31977--31997},
  publisher    = {Association for Computational Linguistics},
  year         = {2026},
  url          = {https://doi.org/10.18653/v1/2026.findings-acl.1600},
  doi          = {10.18653/V1/2026.FINDINGS-ACL.1600},
  bibsource    = {dblp computer science bibliography, https://dblp.org}
}

@misc{Agentic_RAG_R1,
  title       = {Agentic RAG-R1: Enhance Agentic RAG Reasoning Capacity via Reinforcement Learning},
  author      = {Jiang, Xinke and Gao, Jiaran and Qiu, Rihong and Zhang, Zhixin and Zhang, Wentao and Fang, Yue and Ding, Hongxin},
  year        = {2025},
  howpublished= {\url{https://github.com/jiangxinke/Agentic-RAG-R1}},
  note        = {GitHub repository},
}

\appendix

\newpage

\addtocontents{toc}{\protect\setcounter{tocdepth}{2}}

\section*{Appendix}
\tableofcontents

\section{Related Work}
\label{sec:related_work}

\subsection{LLM-Based Multi-Agent Systems}

LLM-based multi-agent systems have emerged as an important paradigm for solving complex tasks ~\cite{DBLP:conf/ijcai/GuoCWCPCW024,chen2025surveyllmbasedmultiagentsystem}. By decomposing a complex problem into sub-tasks and assigning specialized roles to different agents, these systems can improve task planning and collaborative decision making. Representative systems such as AutoGen~\cite{DBLP:journals/corr/abs-2308-08155}, MetaGPT~\cite{DBLP:conf/iclr/HongZCZCWZWYLZR24}, and ChatDev~\cite{DBLP:conf/acl/QianLLCDL0CSCXL24} employ explicit role assignment and structured communication in software development and collaborative decision-making scenarios. Beyond role-based collaboration, decentralized and debate-based systems use dynamic inter-agent interaction to increase reasoning diversity and mitigate single-agent decision bias ~\cite{DBLP:conf/icml/Du00TM24,DBLP:conf/nips/YangCSSQRZ25,cui2025freemad}, although they may introduce redundant communication and complicate global-state consistency. For complex long-horizon tasks, centralized coordination is a widely adopted alternative: a central planner or coordinator performs global task planning, sub-task dispatch, and cross-agent state management. CoAct~\cite{DBLP:journals/corr/abs-2406-13381} develops a global--local hierarchical architecture, MasRouter~\cite{DBLP:conf/acl/YueZLWWCQ25} learns dynamic task routing, and MacNet, OWL, and AFlow improve collaboration structures and workflow optimization ~\cite{DBLP:conf/iclr/QianXW0ZXDDC00025,DBLP:conf/nips/HuZFNYXSJLZWYGL25, DBLP:conf/iclr/ZhangXYTCCZCHWZ25}. STACKPLANNER~\cite{zhang2026stackplanner} further introduces RL-based active memory management, using a Task Memory Stack and a learnable \textsc{Revise} action to retrieve and reuse cross-task experience. Although these studies demonstrate the effectiveness of centralized coordination, they primarily focus on system architecture, communication, or workflow design, with limited attention to learning and optimizing the central agent itself. Our work focuses on this underexplored policy-optimization problem.

\subsection{Reinforcement Learning for Agentic Systems} 

RL has shown substantial potential for improving the complex reasoning capabilities of LLMs~\cite{openaio1}. Representative work such as DeepSeek-R1~\cite{guo2025deepseek} uses Group Relative Policy Optimization (GRPO) to train long-chain reasoning, achieving substantial advances in mathematical and code reasoning~\cite{shao2024deepseekmath}, while DAPO~\cite{DBLP:conf/nips/YuZZYZYDFLLLLLL25} further improves training stability. Motivated by these advances, agentic RL methods learn strategies for tool use and task execution. Search-R1~\cite{jin2025search} trains a model to decide when to invoke a search engine; RAGEN ~\cite{wang2025ragenunderstandingselfevolutionllm} models multi-step agent interaction as an RL problem; and ToolRL ~\cite{qian2025toolrlrewardtoollearning} develops reward mechanisms for tool learning. AEPO~\cite{dong2025agentic} and ARPO~\cite{dong2025agentic2} improve optimization and exploration in high-entropy agentic settings. However, these methods primarily target single-agent settings or homogeneous action spaces, implicitly treating tokens in a sequence as having similar semantic functions and optimization objectives. A central-agent output instead contains two fundamentally different decisions: high-level meta-action selection (Action Tokens) and concrete args generation (Args Tokens), whose gradient statistics and optimization dynamics can differ substantially. Applying a shared policy objective can therefore cause gradient conflicts and policy degradation, motivating our decoupled optimization framework.

\subsection{Functional Specialization in Language Models}

Researchers have increasingly investigated whether LLMs develop internal functional specialization during pretraining. Attention heads, feed-forward network layers, and individual neurons can exhibit specialized responses to particular tasks or types of knowledge~\cite{dai2021knowledge,lape, todd2023function}. Model Shapley~\cite{jiang2025model} quantifies parameter-level importance from a cooperative-game-theoretic perspective, providing a principled basis for functional-partition analysis. At the level of parameter localization, ROME~\cite{meng2022rome} and MEMIT~\cite{meng2023memit} identify specific MLP components associated with factual knowledge, providing evidence of functionally differentiated parameter structures. Diff Pruning~\cite{guo2021diffpruning} and LoRA~\cite{hu2021lora} further constrain updates to compact parameter subspaces, enabling efficient and targeted adaptation. Inspired by these findings, we use forward activation statistics to identify functional partitions associated with Action selection and Args generation, and route their policy gradients to the corresponding subspaces to reduce interference between the two decision types.

\section{Training Configuration}
 \label{Training_Configuration}
 Table~\ref{tab:central_r1_hyperparameters} records the settings explicitly
 available in the current experimental report.
 
 \begin{table*}[ht]
 \centering
 \small
 \setlength{\tabcolsep}{5pt}
 \renewcommand{\arraystretch}{1.1}
 \begin{tabular}{ll}
 \toprule
 \textbf{Hyperparameter} & \textbf{Value} \\
 \midrule
 Rollouts per query $K$ & 4 \\
 GPU configuration & $4\times$ A100-SXM4-80GB (1.5B); $8\times$ A100-SXM4-80GB (3B) \\
 Optimizer & Adam \\
 Learning rate & $2\times10^{-7}$ \\
 Global/per-device batch size & 32 / dynamic (mean 8 for 1.5B; 4 for 3B) \\
 Maximum harness steps & 12 \\
 Maximum input/output length & 16,384 / 1,024 tokens \\
 Sampling temperature / top-$p$ & 0.8 / 1.0 \\
 Clipping coefficient $\epsilon$ & 0.2 \\
 \bottomrule
 \end{tabular}
 \caption{\M~training configuration for the central-capability runs.}
 \label{tab:central_r1_hyperparameters}
 \end{table*}

\section{Computational Resources and Software Environment}

Experiments were conducted on a server with two Intel Xeon Platinum 8488C
processors, providing 96 physical cores and 192 hardware threads, and 503~GB of
system memory. \M~training used NVIDIA A100-SXM4 GPUs with 80~GB of
memory per GPU: four GPUs for Qwen2.5-1.5B-Instruct and eight GPUs for
Qwen2.5-3B-Instruct. The operating system was Ubuntu~22.04.5~LTS. The software
environment used Python~3.11.11, Conda~23.5.2, PyTorch~2.6.0 with CUDA support,
Hugging Face Transformers~4.51.3, and spaCy~3.8.4. Unless otherwise specified,
all experiments used the same hardware and software environment.

\section{Detailed Case Study}
\label{app:case_study}

We present a representative trajectory to illustrate how  \M~separates
high-level dispatch decisions from their args and refines the latter using
sub-agent feedback. As shown in Table~\ref{tab:central_r1_case}, the central
agent preserves the appropriate \textsc{Dispatch} action across two retrieval
rounds while changing its args from broad entity identification to precise
timeline verification. This behavior allows the agent to resolve an ambiguous
temporal clue before returning the correct answer.

\begin{table*}[t]
\centering
\small
\setlength{\tabcolsep}{5pt}
\renewcommand{\arraystretch}{1.05}
\begin{tabular}{
  >{\centering\arraybackslash}m{2.0cm}
  >{\raggedright\arraybackslash}m{12.5cm}}
\hline
\rowcolor{gray!20}
\textbf{Question} & Which Yuri served as USSR president for approximately two
years after heading the KGB for 15 years? \\ \hline
\textbf{Ground Truth} & Yuri Andropov \\ \hline
\textbf{ \M} &

\begin{tcolorbox}[colback=blue!10!white,colframe=white,boxsep=1pt,
left=2pt,right=2pt,top=2pt,bottom=2pt,after=\vspace{-5pt}]
\textbf{Action:} \textsc{Dispatch} $\rightarrow$ Search Agent
\end{tcolorbox}

\begin{tcolorbox}[colback=green!12!white,colframe=white,boxsep=1pt,
left=2pt,right=2pt,top=2pt,bottom=2pt,after=\vspace{-5pt}]
\textbf{Args:} Identify the Yuri who became a Soviet leader after serving as
KGB chairman for approximately 15 years, and verify the relevant timeline.
\end{tcolorbox}

\begin{tcolorbox}[colback=gray!12!white,colframe=white,boxsep=1pt,
left=2pt,right=2pt,top=2pt,bottom=2pt,after=\vspace{-5pt}]
\textbf{Observation:} Yuri Vladimirovich Andropov chaired the KGB from 1967 to
1982 and subsequently became the Soviet leader; the exact duration of his later
tenure remains uncertain.
\end{tcolorbox}

\begin{tcolorbox}[colback=violet!10!white,colframe=white,boxsep=1pt,
left=2pt,right=2pt,top=2pt,bottom=2pt,after=\vspace{-5pt}]
\textbf{Reason:} Andropov served as General Secretary for roughly 15 months,
whereas the question says ``approximately two years.'' The dates and possible
alternative candidates should therefore be verified before answering.
\end{tcolorbox}

\begin{tcolorbox}[colback=blue!10!white,colframe=white,boxsep=1pt,
left=2pt,right=2pt,top=2pt,bottom=2pt,after=\vspace{-5pt}]
\textbf{Action:} \textsc{Dispatch} $\rightarrow$ Search Agent
\end{tcolorbox}

\begin{tcolorbox}[colback=green!12!white,colframe=white,boxsep=1pt,
left=2pt,right=2pt,top=2pt,bottom=2pt,after=\vspace{-5pt}]
\textbf{Args:} Verify Andropov's exact tenure dates and determine whether any
other person named Yuri satisfies both the KGB-service and presidency clues.
\end{tcolorbox}

\begin{tcolorbox}[colback=gray!12!white,colframe=white,boxsep=1pt,
left=2pt,right=2pt,top=2pt,bottom=2pt,after=\vspace{-5pt}]
\textbf{Observation:} The evidence confirms Andropov as the unique candidate;
the phrase ``approximately two years'' is an approximate description of his
tenure.
\end{tcolorbox}

\begin{tcolorbox}[colback=yellow!15!white,colframe=white,boxsep=1pt,
left=2pt,right=2pt,top=2pt,bottom=2pt]
\textbf{Final Answer:} Yuri Andropov
\end{tcolorbox}
\\ \hline
\end{tabular}
\caption{Case study of  \M~on an ambiguous multi-hop question. Blue and
green boxes denote Actions and Args, respectively; gray boxes show sub-agent
observations, violet shows central-agent reasoning, and yellow shows the final
answer.}
\label{tab:central_r1_case}
\end{table*}


\section{Use of Large Language Models}
\label{app:llm_use}

Large language models were used only as auxiliary tools for language refinement and programming-related assistance. They were used to improve grammatical accuracy, readability, and presentation, and to provide general coding or debugging suggestions. All model-assisted outputs were reviewed and validated by the authors before inclusion. The authors were responsible for the research conception, methodological design, experimental execution, interpretation of results, and final conclusions.

\end{document}